\documentclass[letterpaper,journal,10pt]{IEEEtran}

\usepackage{cite}

\usepackage{graphicx}

\usepackage[cmex10]{amsmath}
\usepackage{amssymb}
\usepackage{fixltx2e}

\usepackage{multirow}
\usepackage{booktabs}
\usepackage{dcolumn}

\usepackage{ifpdf}
\ifpdf
\usepackage[draft,pdfborder={0 0 0}]{hyperref}
\fi

\usepackage{bm}
\usepackage[varg]{txfonts}
\let\mathbb=\varmathbb
\DeclareSymbolFont{letters}{OML}{ztmcm}{m}{it}

\usepackage{tikz,pgf}
\usepackage{pgfplots}
\usetikzlibrary{shapes,positioning,arrows,decorations.markings,fit,calc,patterns}

\usepackage{subfig}

\usepackage{color}

\DeclareMathOperator*{\argmin}{arg\,min}

\title{Fast Polar Decoders: Algorithm and Implementation}

\author{\IEEEauthorblockN{Gabi Sarkis\IEEEauthorrefmark{1}, Pascal Giard\IEEEauthorrefmark{1}\IEEEauthorrefmark{2}, Alexander Vardy\IEEEauthorrefmark{3},
    Claude Thibeault\IEEEauthorrefmark{2}, and Warren J. Gross\IEEEauthorrefmark{1}}\\
\IEEEauthorblockA{\small\IEEEauthorrefmark{1}Department of Electrical and Computer Engineering, McGill University, Montr\'eal, Qu\'ebec, Canada\\
  Email: gabi.sarkis@mail.mcgill.ca, pascal.giard@mail.mcgill.ca, and warren.gross@mcgill.ca}\\
\IEEEauthorblockA{\small\IEEEauthorrefmark{2}Department of Electrical Engineering, \'Ecole de technologie sup\'erieure, Montr\'eal, Qu\'ebec, Canada\\
  Email: claude.thibeault@etsmtl.ca}\\
\IEEEauthorblockA{\small\IEEEauthorrefmark{3}University of California San Diego, La Jolla, CA. USA\\
  Email: avardy@ucsd.edu}
\\
}

\ifpdf
\hypersetup{pdfauthor={Gabi Sarkis, Pascal Giard, Alexander Vardy, Claude Thibeault, and Warren J. Gross}%
           ,pdftitle={Fast Polar Decoders: Algorithm and Implementation}}%
\fi

\newcommand{\est}[1]{\hat{u}_{#1}}
\newcommand{\estx}[1]{\hat{x}_{#1}}
\newcommand{\esta}[1]{\hat{a}_{#1}}
\newcommand{\estvec}[2]{\hat{\mvec{u}}_{#1}^{#2}}

\newcommand{\llr}[1]{\lambda_{#1}}

\newcommand{\nz}{\mathcal{N}^0}
\newcommand{\no}{\mathcal{N}^1}
\newcommand{\nr}{\mathcal{N}^R}
\newcommand{\nml}{\mathcal{N}^{\text{ML}}}
\newcommand{\nrep}{\mathcal{N}^{\text{REP}}}
\newcommand{\nspc}{\mathcal{N}^{\text{SPC}}}
\newcommand{\nrepspc}{\mathcal{N}^{\text{REP-SPC}}}

\newcommand{\code}{\mathcal{C}}

\newcommand{\sgn}[1]{\text{sign}(#1)}
\newcommand{\mvec}[1]{\bm{#1}}
\newcommand{\prob}[1]{\text{Pr}[#1]}

\begin{document}

\maketitle

\begin{abstract}
Polar codes provably achieve the symmetric capacity of a memoryless channel while having an explicit construction. The adoption of polar codes however, has been hampered by the low throughput of their decoding algorithm. This work aims to increase the throughput of polar decoding hardware by an order of magnitude relative to successive-cancellation decoders and is more than 8 times faster than the current fastest polar decoder. We present an algorithm, architecture, and FPGA implementation of a flexible, gigabit-per-second polar decoder.
\end{abstract}

\section{Introduction}
\label{sec:introduction}

Polar codes \cite{Arikan2009} are the first error-correcting codes with an explicit construction to provably achieve the symmetric capacity of memoryless channels. They have two properties that are of interest to data storage systems: a very low error-floor due to their large stopping distance \cite{Eslami2010}, and low-complexity implementations \cite{Leroux2013}. However, polar codes have two drawbacks: their performance at short to moderate lengths is inferior to that of other codes, such as low-density parity-check (LDPC) codes; and their low-complexity decoding algorithm, successive-cancellation (SC), is serial in nature, leading to low decoding throughput \cite{Leroux2013}.

Multiple methods exist to improve the error-correction performance of polar codes. Using list, and list-CRC decoding \cite{Tal2012} improves performance significantly. Alternatively, one can increase the length of the polar code. Using a code length corresponding to the block length of current hard drives \cite{chicoine2007}, as we show in this work, results in a polar decoder with lower complexity than an LDPC decoder with similar error-correction performance and the same rate. Specifically, a (32768, 27568) polar code has slightly worse error-correction performance than the (2048, 1723) LDPC code of the 10GBASE-T (802.3an) standard in the low signal-to-noise ratio (SNR) region but better performance at frame error rates (FER) lower than $7\times 10^{-7}$, with the high SNR region being more important for storage systems. In addition, polar codes can be made to perform better than the LDPC code starting at FER of $2\times 10^{-4}$ as shown in Section~\ref{sec:polar-codes}.

Among the many throughput-improving methods proposed in literature, simplified successive-cancellation (SSC) \cite{Alamdar-Yazdi2011} and simplified successive-cancellation with maximum-likelihood nodes (ML-SSC) \cite{Sarkis2013} offer the largest improvement over SC decoding. This throughput increase is achieved by exploiting the recursive nature of polar codes---where every polar code of length $N$ is formed from two constituent polar codes of length $N/2$---and decoding the constituent codes directly, without recursion, when possible. SSC decodes constituent codes of rates 0 and 1 directly and ML-SSC additionally enables the direct decoding of smaller constituent codes.

In this work, we focus on improving the throughput of polar decoders. By building on the ideas used for SSC and ML-SSC decoding, namely decoding constituent codes without recursion, and recognizing further classes of constituent codes that can be directly decoded, we present a polar decoder that, for a (32768, 29492) code, is 40 times faster than the best SC decoder \cite{Leroux2013} when implemented on the same field-programmable gate-array (FPGA). For a (16384, 14746) code, our decoder is more than 8 times faster than the state of the art polar decoder in litterature \cite{Pamuk2013}, again when implemented on the same FPGA. Additionally, the proposed decoder is flexible and can decode any polar code of a given length.

We start this paper by reviewing polar codes, their construction, and the successive-cancellation decoding algorithm in Section~\ref{sec:polar-codes}. The SSC and ML-SSC decoding algorithms are reviewed in Section~\ref{sec:ssc-mlssc}. We present our improved decoding algorithm in Section~\ref{sec:algorithm}, including new constituent code decoders. The decoder architecture is discussed in detail in sections~\ref{sec:architecture:top}, \ref{sec:architecture:data}, and \ref{sec:architecture:processor}. Implementation results, showing that the proposed decoder has lower complexity on an FPGA than the 10GBASE-T LDPC decoder with the same rate and comparable error-correction performance, are presented in Section~\ref{sec:implementation}.

We focus on two codes: a (32768, 29492) that has a rate of 0.9 making it suitable for storage systems; and a (32768, 27568) that is comparable to the popular 10GBASE-T LDPC code in error-correction performance and has the same rate, which enables the implementation complexity comparison in Section~\ref{sec:implementation}.

\section{Polar Codes}
\label{sec:polar-codes}
\subsection{Construction of Polar Codes}
\label{sec:polar-codes:construction}
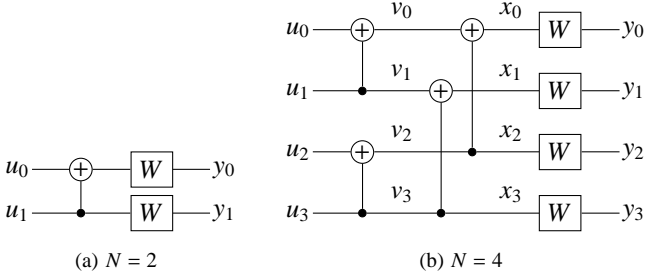
\begin{figure}[t]
  \centering
  \subfloat[$N=2$]{\label{fig:pc2}\newcommand{\ubit}[1]{$u_{#1}$}
\newcommand{\fbit}[1]{\color{gray}$u_{#1}$}
\begin{tikzpicture}

\usetikzlibrary{shapes,positioning,arrows,decorations.markings,fit}

\definecolor{varnode_fill}{RGB}{0,0,0}
\definecolor{chknode_fill}{RGB}{255,255,255}

\tikzset{
  chknode/.style={draw,fill=chknode_fill,circle,minimum size=0.3cm, inner sep=0},
  varnode/.style={draw,fill=varnode_fill,circle,minimum size=0.1cm, inner sep=0},
  channel/.style={draw,fill=white,rectangle},
  sep/.style={rectangle,minimum width=0.3cm, inner sep=0},
  bit/.style={circle, inner sep = 0}
}

\matrix[row sep=1mm, column sep=1mm] {
 	\node[bit] (n0s0) {\ubit{0}}; & \node[sep] {}; & \node[chknode] (n0s1) {$+$}; &  \node[sep] {}; & \node[channel] (n0s2) {$W$}; & \node[sep] {}; & \node[bit] (n0s3) {$y_0$};\\
 	\node[bit] (n1s0) {\ubit{1}}; & \node[sep] {}; & \node[varnode] (n1s1) {}; &  \node[sep] {}; & \node[channel] (n1s2) {$W$}; & \node[sep] {}; & \node[bit] (n1s3) {$y_1$};\\
};

\path[-] (n0s0) edge (n0s1);
\path[-] (n0s1) edge (n0s2);
\path[-] (n0s2) edge (n0s3);

\path[-] (n1s0) edge (n1s1);
\path[-] (n1s1) edge (n1s2);
\path[-] (n1s2) edge (n1s3);

\path[-] (n1s1) edge (n0s1);

\end{tikzpicture}}
  \subfloat[$N=4$]{\label{fig:pc4}\newcommand{\ubit}[1]{$u_{#1}$}
\newcommand{\fbit}[1]{\color{gray}$u_{#1}$}
\begin{tikzpicture}

\usetikzlibrary{shapes,positioning,arrows,decorations.markings,fit}

\definecolor{varnode_fill}{RGB}{0,0,0}
\definecolor{chknode_fill}{RGB}{255,255,255}

\tikzset{
  chknode/.style={draw,fill=chknode_fill,circle,minimum size=0.3cm, inner sep=0},
  varnode/.style={draw,fill=varnode_fill,circle,minimum size=0.1cm, inner sep=0},
  channel/.style={draw,fill=white,rectangle},
  sep/.style={rectangle,minimum width=0.31cm, inner sep=0},
  bit/.style={circle, inner sep = 0}
}

\tikzset{blue dotted/.style={draw=blue!50!white, line width=1pt,
    dash pattern=on 4pt off 4pt,
    inner sep=0.5mm, rectangle, rounded corners}};

\tikzset{blue dotted tight/.style={draw=blue!50!white, line width=1pt,
    dash pattern=on 4pt off 4pt,
    inner sep=0mm, rectangle, rounded corners}};

\matrix[row sep=1mm, column sep=1mm] {
  \node[bit] (n0u) {\ubit{0}}; & \node[sep] {}; & \node[chknode] (n0s1) {$+$}; & \node[sep,label=$v_0$] {}; && \node[chknode] (n0s2) {$+$}; &  \node[sep,label={$x_0$}] {}; & \node[channel] (n0c) {$W$}; & \node[sep] {}; & \node[bit] (n0y) {$y_0$};\\
  \node[bit] (n1u) {\ubit{1}}; & \node[sep] {}; & \node[varnode] (n1s1) {};    & \node[sep,label=$v_1$] {}; &  \node[chknode] (n1s2) {$+$}; && \node[sep,label={$x_1$}] {}; & \node[channel] (n1c) {$W$}; & \node[sep] {}; & \node[bit] (n1y) {$y_1$};\\
  \node[bit] (n2u) {\ubit{2}}; & \node[sep] {}; & \node[chknode] (n2s1) {$+$}; & \node[sep,label=$v_2$] {}; && \node[varnode] (n2s2) {};    &  \node[sep,label={$x_2$}] {}; & \node[channel] (n2c) {$W$}; & \node[sep] {}; & \node[bit] (n2y) {$y_2$};\\
  \node[bit] (n3u) {\ubit{3}}; & \node[sep] {}; & \node[varnode] (n3s1) {};    & \node[sep,label=$v_3$] {}; &  \node[varnode] (n3s2) {};    && \node[sep,label={$x_3$}] {}; & \node[channel] (n3c) {$W$}; & \node[sep] {}; & \node[bit] (n3y) {$y_3$};\\
};

\path[-] (n0u) edge (n0s1);
\path[-] (n0s1) edge (n0s2);
\path[-] (n0s2) edge (n0c);
\path[-] (n0c) edge (n0y);

\path[-] (n1u) edge (n1s1);
\path[-] (n1s1) edge (n1s2);
\path[-] (n1s2) edge (n1c);
\path[-] (n1c) edge (n1y);

\path[-] (n1s1) edge (n0s1);

\path[-] (n2u) edge (n2s1);
\path[-] (n2s1) edge (n2s2);
\path[-] (n2s2) edge (n2c);
\path[-] (n2c) edge (n2y);

\path[-] (n3u) edge (n3s1);
\path[-] (n3s1) edge (n3s2);
\path[-] (n3s2) edge (n3c);
\path[-] (n3c) edge (n3y);

\path[-] (n3s1) edge (n2s1);

\path[-] (n3s2) edge (n1s2);
\path[-] (n2s2) edge (n0s2);

\end{tikzpicture}}
  \caption{Construction of polar codes of lengths 2 and 4}
\end{figure}

By exploiting channel polarization, polar codes approach the symmetric capacity of a channel as the code length, $N$, increases. The polarizing construction when $N = 2$ is shown in Fig.~\ref{fig:pc2}, where the probability of correctly estimating bit $u_0$ decreases; while that of bit $u_1$ increases compared to when the bits are transmitted without any transformation over the channel $W$. Channels can be combined recursively to create longer codes, Fig.~\ref{fig:pc4} shows the case of $N = 4$. As $N \to \infty$, the probability of successfully estimating each bit approaches 1 (perfectly reliable) or 0.5 (completely unreliable), and the proportion of reliable bits approaches the symmetric capacity of $W$ \cite{Arikan2009}.

To create an ($N$, $k$) polar code, $N$ copies of the channel $W$ are transformed using the polarizing transform and the $k$ most reliable bits, called the information bits, are used to send information bits; while the $N - k$ least reliable bits, called the frozen bits, are set to 0. Determining the locations of the information and frozen bits depends on the type and conditions of $W$ and is investigated in detail in \cite{Tal2011a}. Therefore, a polar code is constructed for a given channel and channel condition.
A polar code of length $N$ can be represented using a generator matrix, $G_N = F_N = F_2^{\otimes \log_2 N}$, where $F_2 = \left[ \begin{smallmatrix} 1 & 0 \\ 1 & 1\end{smallmatrix} \right]$ and $^{\otimes}$ is the Kronecker power.
The frozen bits are indicated by setting their values to 0 in the source vector $\mvec{u}$.

Polar codes can be encoded systematically to improve bit error-rate (BER)\cite{Arikan2011}. Furthermore, systematic polar codes are a natural fit for the SSC and ML-SSC algorithms \cite{Sarkis2013}.

In \cite{Arikan2009}, bit-reversed indexing is used, which changes the generator matrix by multiplying it with a bit-reversal operator $B$, so that $G = B F$.
In this work, we use natural indexing to review and introduce algorithms for reasons of clarity. However, it was shown in \cite{Leroux2013} that bit-reversed indexing significantly reduced data-routing complexity in a hardware implementation; therefore, we used it to implement our decoder architecture. In Section~\ref{sec:sys-bit-rev}, we review how to combine systematic encoding and bit-reversal without using any interleavers.

\subsection{Successive-Cancellation Decoding}
\label{sec:polar-codes:sc}
Polar codes achieve the channel capacity asymptotically in code length when decoded using the successive-cancellation (SC) decoding algorithm, which sequentially estimates the bits $\est{i}$, where $0 \leq i < N$, using the channel output $\mvec{y}$ and the previously estimated bits, $\est{0}$ to $\est{i - 1}$, denoted $\estvec{0}{i - 1}$, according to:
\begin{equation}
\label{eq:sc:spa:decision}
  \est{i} = \begin{cases}
    0,& \text{ if } \llr{u_i} \geq 0;\\
    1,& \text{ otherwise}.
  \end{cases}
\end{equation}
Where $\llr{u_i}$ is the log-likelihood ratio (LLR) defined as $\prob{\mvec{y}, \estvec{0}{i-1} | \est{i} = 0}/\prob{\mvec{y}, \estvec{0}{i-1} | \est{i} = 1}$ and can be calculated recursively using the min-sum (MS) approximation according to \cite{Leroux2013}
\begin{equation}
\label{eq:sc:ms:f}
\llr{u_0} = f(\llr{v_0}, \llr{v_1}) = \sgn{\llr{v_0}}\sgn{\llr{v_1}} \min(|\llr{v_0}|, |\llr{v_1}|);
\end{equation}
and
\begin{equation}
\label{eq:sc:ms:g}
\llr{u_1} = g(\llr{v_0}, \llr{v_1}, \est{0}) = \begin{cases}
\llr{v_0} + \llr{v_1} & \text{when } \est{0} = 0,\\
-\llr{v_0} + \llr{v_1} & \text{when } \est{0} = 1.\\
\end{cases}
\end{equation}

\subsection{Performance of SC Decoding}
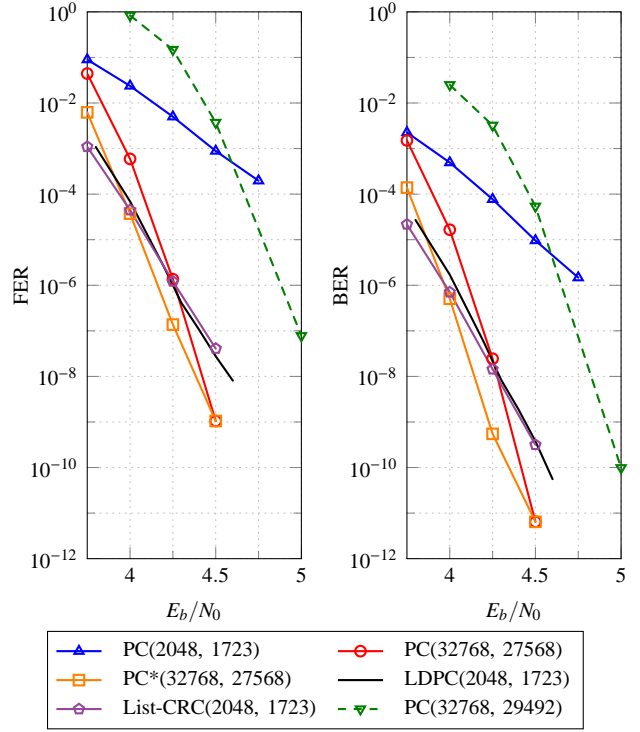
\begin{figure}[t]
\centering
\centering
\usetikzlibrary{plotmarks}

\definecolor{darkgreen}{RGB}{0, 128, 0}
\definecolor{mypurple}{RGB}{153, 71, 155}

\begin{tikzpicture}

 \begin{semilogyaxis}
    [
    width=0.5\columnwidth,
    height=\columnwidth,
    line cap=round,
    every axis/.append style={font=\footnotesize},
    every axis y label/.style={at={(ticklabel cs:0.5)},rotate=90},
    every axis x label/.style={at={(ticklabel cs:0.5)},anchor=near ticklabel},
    every axis plot/.append style={thick},
    xmin=3.75,
    xmax=5,
    xlabel={$E_b/N_0$},
    xtick={4, 4.5, 5},
    minor x tick num={1},
    ymax=1.0,
    ymin=1.0e-12,
    ytickten={0,-2,...,-12},
    minor ytick={1e-1, 1e-3, 1e-5, 1e-7, 1e-9, 1e-11},
    ylabel={FER},
    legend columns = 2,
    legend style={
      anchor={center},
      cells={anchor=west},
      column sep= 2mm,
      font=\footnotesize,
    },
    legend to name=perf-legend,
    grid=both,
    grid style=dotted,
    ]
    
     \addplot[color=blue, mark=triangle] table[x=snr_db,y=FER] {data/pc-ssc2k.o7728};
    \addlegendentry{PC(2048, 1723)}

    \addplot[color=red, mark=o] table[x=snr_db,y=FER] {data/pc-ssc32k.o7727};
    \addlegendentry{PC(32768, 27568)}

    \addplot[color=orange, mark=square] table[x=snr_db,y=FER] {data/pc32kopt.txt};
    \addlegendentry{PC*(32768, 27568)}

    \addplot[color=black] table[x=snr_db,y=FER] {data/10gige.dat};
    \addlegendentry{LDPC(2048, 1723)}

    \addplot[color=mypurple, mark=pentagon] table[x=snr_db,y=FER] {data/pc-a2k.o7724};
    \addlegendentry{List-CRC(2048, 1723)}

    \addplot[color=darkgreen, dashed, mark=triangle, mark options={solid, rotate=60}] table[x=snr_db,y=FER] {data/mlssc32k.o8134};
    \addlegendentry{PC(32768, 29492)}

  \end{semilogyaxis}
\end{tikzpicture}
\begin{tikzpicture}

  \begin{semilogyaxis}
    [
    width=0.5\columnwidth,
    height=\columnwidth,
    line cap=round,
    every axis/.append style={font=\footnotesize},
    every axis y label/.style={at={(ticklabel cs:0.5)},rotate=90},
    every axis x label/.style={at={(ticklabel cs:0.5)},anchor=near ticklabel},
    every axis plot/.append style={thick},
    xmin=3.75,
    xmax=5,
    xlabel={$E_b/N_0$},
    xtick={4, 4.5, 5},
    minor x tick num={1},
    ymax=1.0,
    ymin=1.0e-12,
    ytickten={0,-2,...,-12},
    minor ytick={1e-1, 1e-3, 1e-5, 1e-7, 1e-9, 1e-11},
    ylabel={BER},
    grid=both,
    grid style=dotted,
    ]
    
    \addplot[color=blue, mark=triangle] table[x=snr_db,y=BER] {data/pc-ssc2k.o7728};

    \addplot[color=red, mark=o] table[x=snr_db,y=BER] {data/pc-ssc32k.o7727};

    \addplot[color=orange, mark=square] table[x=snr_db,y=BER] {data/pc32kopt.txt};

    \addplot[color=black] table[x=snr_db,y=BER] {data/10gige.dat};

    \addplot[color=mypurple, mark=pentagon] table[x=snr_db,y=BER] {data/pc-a2k.o7724};

    \addplot[color=darkgreen, dashed, mark=triangle, mark options={solid, rotate=60}] table[x=snr_db,y=BER] {data/mlssc32k.o8134};

  \end{semilogyaxis}

\end{tikzpicture}
\\
\ref{perf-legend}
\caption{Error-correction performance of polar codes compared with that of an LDPC code with the same rate. In addition to the performance of a rate 0.9 polar code.}
\label{fig:pc:perf}
\end{figure}

Fig.~\ref{fig:pc:perf} shows the error-correction performance of the (2048, 1723) 10GBASE-T LDPC code compared to that of polar codes of the same rate. These results were obtained for the binary-input additive white Gaussian noise (AWGN) channel with random codewords and binary phase-shift keying (BPSK) modulation. The first observation to be made is that the performance of the (2048, 1723) polar code is significantly worse than that of the LDPC code. The polar code of length 32768, labeled PC(32768, 27568), was constructed to be optimal for $E_b/N_0 = 4.5$ dB and performs worse than the LDPC code until the $E_b/N_0 = 4.25$ dB. Past that point, it outperforms the LDPC code with a growing gap. The last polar error-rate curve, labeled PC*(32768, 27568), combines the results of two (32768, 27568) polar codes. One is constructed for 4.25 dB and used up to that point, and the other is constructed for 4.5 dB. Due to the regular structure of polar codes, it is simple to build a decoder that can decode any polar code of a given length. Therefore, it is simpler to change polar codes in a system than it is to change LDPC codes.

From these results, it can be concluded that a (32768, 27568) polar code constructed for 4.5 dB or a higher $E_b/N_0$ is required to outperform the (2048, 1723) LDPC one in the low error-rate region, and a combination of different polar codes can be used to outperform the LDPC code even in high error-rate regions. Even though the polar code has a longer length, its decoder still has a lower implementation complexity than the LDPC decoder as will be shown in Section~\ref{sec:implementation}.

Decoding the (2048, 1723) code using the list-CRC algorithm \cite{Tal2012}, with a list size of 32 and a 32-bit CRC, reduces the gap with the LDPC code to the point where the two codes have similar performance as shown in Fig.~\ref{fig:pc:perf}. However, in spite of this improvement, we do not discuss list-CRC decoding in this work as it cannot directly accommodate the proposed throughput-improving techniques, which are designed to provide a single estimate instead of a list of potential candidates. Further research is required to adapt some of these techniques to list decoding.

The throughput of SC decoding is limited by its serial nature: the fastest implementation currently is an ASIC decoder for a (1024, 512) polar code with an information throughput of 48.75~Mbps when running at 150~MHz \cite{Mishra2012}; while the fastest decoder for a code of length 32768 is FPGA-based and has a throughput of 26~Mbps for the (32768, 27568) code \cite{Leroux2013}. This low throughput renders SC decoders impractical for most systems; however, it can be improved significantly by using the SSC or the ML-SSC decoding algorithms.

\section{SSC and ML-SSC Decoding}
\label{sec:ssc-mlssc}
\subsection{Tree Structure of an SC Decoder}
A polar code of length $N$ is the concatenation of two polar codes of length $N/2$. Since this construction is recursive, as mentioned in Section~\ref{sec:polar-codes:construction}, a binary tree is a natural representation for a polar code where each node corresponds to a constituent code.  The tree representation is presented in detail in \cite{Alamdar-Yazdi2011} and \cite{Sarkis2013}. Fig.~\ref{fig:sc-tree} shows the tree representation for an (8, 3) polar code where the white and black leaves correspond to frozen and information bits, respectively.

A node $v$, corresponding to a constituent code of length $N_v$, receives a real-valued message vector, $\mvec{\alpha}_v$, containing the soft-valued input to the constituent polar decoder, from its parent node. It calculates the soft-valued input to its left child, $\mvec{\alpha}_l$ using \eqref{eq:sc:ms:f}. Once the constituent codeword estimate, $\mvec{\beta}_l$, from the left child is ready, it is used to calculate the input to the right, $\mvec{\alpha}_r$, according to \eqref{eq:sc:ms:g}. Finally, $\mvec{\beta}_v$ is calculated from $\mvec{\beta}_l$ and $\mvec{\beta}_r$ as
\begin{equation}
\label{eq:xor}
\mvec{\beta}_v[i] = \begin{cases}
\mvec{\beta}_l[i] \oplus \mvec{\beta}_r[i], & \text{when } i < N_v/2;\\
\mvec{\beta}_r[i - N_v/2], & \text{otherwise.}
\end{cases}
\end{equation}

For leaf-nodes, $\mvec{\beta}_v$ is 0 if the node is frozen. Otherwise, it is calculated using threshold detection, defined for an LLR-based decoder as:
\begin{equation}
\label{eq:threshold}
\mvec{\beta}_v = \begin{cases}
0, & \text{when } \mvec{\alpha}_v \geq 0;\\
1, & \text{otherwise.}
\end{cases}
\end{equation}

The input to the root node is the LLR values calculated from the channel output, and its output is the estimated systematic codeword.

\begin{figure}[t]
  \centering
  \subfloat[SC]{\label{fig:sc-tree}\begin{tikzpicture}[baseline = (0_7.center),
        level/.style={level distance = 6mm},
        level 1/.style={sibling distance=19mm, edge from parent/.style={draw,black,line width=2pt}},
        level 2/.style={level distance=10mm, sibling distance=9.5mm, edge from parent/.style={draw,black,line width=1pt}},
        level 3/.style={sibling distance=4.7mm, edge from parent/.style={draw,black,line width=0.5pt}},
        ]

\tikzset{
frozen/.style={thick,draw=black,fill=white,minimum size=3mm,circle, inner sep=0},
fullspace/.style={thick,draw=black,fill=black,minimum size=3mm,circle, inner sep = 0},
mixed/.style={thick,draw=black,fill=gray,minimum size=3mm,circle, inner sep = 0},
ml_mixed/.style={thick,draw=black,fill=blue,minimum size=3mm,circle, inner sep = 0}
}

\node[mixed] (p){} [grow=left]
	child {node[frozen] (2_0){}
		child {node[frozen] (1_0){}
			child {node[frozen] (a0_0){}
			}
			child {node[frozen] (a0_1){}
			}
		}
		child {node[frozen] (1_2){}
			child {node[frozen] (0_2){}
			}
			child {node[frozen] (0_3){}
			}
		}
	}
	child {node[mixed] (v){$v$}
		child {node[mixed, label={[label distance = -1mm]left}] (cl){}
			child {node[frozen] (0_4){}
			}
			child {node[fullspace] (0_5){}
			}
		}
		child {node[fullspace, label={[label distance = -1mm]below:right}] (cr){}
			child {node[fullspace] (0_6){}
			}
			child {node[fullspace] (0_7){}
			}
		}
	}
;

\draw[<-] ($(v.north east) - (1mm, -1mm)$) -- node[above left=-2mm] {\footnotesize $\alpha_v$} ($(p.south west) - (1mm, 0mm)$);
\draw[<-] ($(p.south west) + (1mm, -1mm)$) -- node[below right=-2mm] {\footnotesize $\beta_v$} ($(v.north east) + (1mm, 0mm)$);

\draw[<-] ($(v.north west) + (-1.6mm, -1mm)$) -- node[left=1mm] {\footnotesize $\beta_l$} ($(cl.south east) + (0mm, -0.6mm)$);
\draw[<-] ($(cl.south east) - (-1.6mm, -1mm)$) -- node[above right=-2mm] {\footnotesize$\alpha_l$} ($(v.north west) - (0mm, -0.6mm)$);

\draw[<-] ($(v.south west) + (0mm, -0.6mm)$) -- node[below right=-2mm] {\footnotesize $\beta_r$} ($(cr.north east) + (1.6mm, -1mm)$);
\draw[<-] ($(cr.north east) - (0mm, -0.6mm)$) -- node[left=1mm] {\footnotesize $\alpha_r$} ($(v.south west) - (1.6mm, -1mm)$);

\end{tikzpicture}}
  \quad
  \subfloat[SSC]{\label{fig:ssc-tree} \begin{tikzpicture}[baseline=(base.center),
        level/.style={level distance = 6mm},
        level 1/.style={sibling distance=19mm, edge from parent/.style={draw,black,line width=2pt}},
        level 2/.style={sibling distance=9mm, edge from parent/.style={draw,black,line width=1pt}},
        level 3/.style={sibling distance=4mm, edge from parent/.style={draw,black,line width=0.5pt}},
        ]

\tikzset{
frozen/.style={thick,draw=black,fill=white,minimum size=3mm,circle, inner sep=0},
fullspace/.style={thick,draw=black,fill=black,minimum size=3mm,circle, inner sep = 0},
mixed/.style={thick,draw=black,fill=gray,minimum size=3mm,circle, inner sep = 0},
ml_mixed/.style={thick,draw=black,fill=blue,minimum size=3mm,circle, inner sep = 0}
}

\node[mixed] (3_0){} [grow=left]
	child {node[frozen] (2_0){}
	}
	child {node[mixed] (2_1){}
		child {node[mixed] (1_2){}
			child {node[frozen] (0_4){}
			}
			child {node[fullspace] (0_5){}
			}
		}
		child {node[fullspace] (1_3){}
		}
	}
;

\node [circle,below= 0.27mm of 1_3.base] (base) {};

\end{tikzpicture}}
  \quad
  \subfloat[ML-SSC]{\label{fig:ml-ssc-tree} \begin{tikzpicture}[baseline=(base),
        level/.style={level distance = 6mm},
        level 1/.style={sibling distance=19mm, edge from parent/.style={draw,black,line width=2pt}},
        level 2/.style={sibling distance=9mm, edge from parent/.style={draw,black,line width=1pt}},
        level 3/.style={sibling distance=4mm, edge from parent/.style={draw,black,line width=0.5pt}},
        ]

\tikzset{
frozen/.style={thick,draw=black,fill=white,minimum size=3mm,circle, inner sep=0},
fullspace/.style={thick,draw=black,fill=black,minimum size=3mm,circle, inner sep = 0},
mixed/.style={thick,draw=black,fill=gray,minimum size=3mm,circle, inner sep = 0},
ml_mixed/.style={thick,draw=black,pattern=north east lines,pattern color=blue,minimum size=3mm,circle, inner sep = 0}
}

\node[mixed] (3_0){} [grow=left]
	child {node[frozen] (2_0){}
	}
	child {node[mixed] (2_1){}
                child {node[ml_mixed] (3_1) {}}
                child {node[fullspace] (3_2) {}}
	}
;

\node [circle,below= 0.27mm of 3_2.base] (base) {};
\node[circle, left= 6mm of 2_1] (pad1) {};
\node[circle, right= 5mm of 3_0] (pad2) {};
\end{tikzpicture}}
  \caption{Decoder trees corresponding to the SC, SSC, and ML-SSC decoding algorithms}
\end{figure}

\subsection{SSC and ML-SSC Decoder Trees}
In \cite{Alamdar-Yazdi2011}, it was observed that a tree with only frozen leaf-nodes rooted in a node $\nz$, does not need to be traversed as its output will always be a zero-vector. Similarly, it was shown that the output of a tree with only information leaf-nodes rooted in $\no$ can be obtained directly by performing threshold detection \eqref{eq:threshold} on the soft-information vector $\mvec{\alpha_v}$, without any additional calculations. Therefore, the decoder tree can be pruned reducing the number of node visitations and latency. The remaining nodes, denoted $\nr$ as they correspond to codes of rate $0 < R< 1$, perform their calculations as in the SC decoder. The pruned tree for an SSC decoder is shown in Fig.~\ref{fig:ssc-tree} and requires nine time steps compared to the 14 steps required to traverse the SC tree in Fig.~\ref{fig:sc-tree}.

ML-SSC further prunes the decoder tree by using exhaustive-search maximum-likelihood (ML) decoding to decode any constituent code, $\code$, while meeting resource constraints \cite{Sarkis2013}. 
The (8, 3) polar decoder utilizing these $\nml$ nodes, and whose tree is shown in Fig.~\ref{fig:ml-ssc-tree}, where $\nml$ is indicated with a striped pattern and is constrained to $N_v = 2$, requires 7 time steps to estimate a codeword.

\subsection{Performance}
In \cite{Sarkis2013}, it was shown that under resource constraints the information throughput of SSC and ML-SSC decoding increases faster than linearly as the code rate increases, and approximately logarithmically as the code length increases. For example, it was estimated that for a rate 0.9 polar code of length 32768, which is constructed for $E_b/N_0 = 3.47$ dB, the information throughput of a decoder running at 100 MHz using SC decoding is $\sim 45$ Mbit/s and increases by 20 times to 910 Mbit/s when using ML-SSC decoding.
The throughput of SSC and ML-SSC is affected by the code construction parameters as they affect the location of frozen bits, which in turn affects the tree structure of the decoder and the number of nodes that can be directly decoded. For example, constructing the rate 0.9, length 32768 polar code for an $E_b/N_0$ of 5.0 dB instead of 3.47 dB, reduces the information throughput of the decoder to 520 Mbit/s assuming the same clock frequency of 100 MHz. While this is a significant reduction, the decoder remains 11 times faster than an SC decoder.

It was noted in \cite{Sarkis2013} that the error-correction performance of polar codes is not tangibly altered by the use of the SSC or ML-SSC decoding algorithms.

\subsection{Systematic Encoding and Bit-Reversal}
\label{sec:sys-bit-rev}
In \cite{Arikan2011}, it was stated that systematic encoding and bit-reversed indexing can be combined. In this section, we review how the information bits can be presented at the output of the decoder in the order in which they were presented by the source, without the use of interleavers. This is of importance to the SSC decoding algorithm as it presents its output in parallel and would otherwise require an $N$ bit parallel interleaver of significant complexity.
The problem is compounded in a resource-constrained, semi-parallel SSC decoder that stores its output one word at a time in memory: since two consecutive information bits might not be in the same memory word, memory words will be visited multiple times, significantly increasing decoding latency.

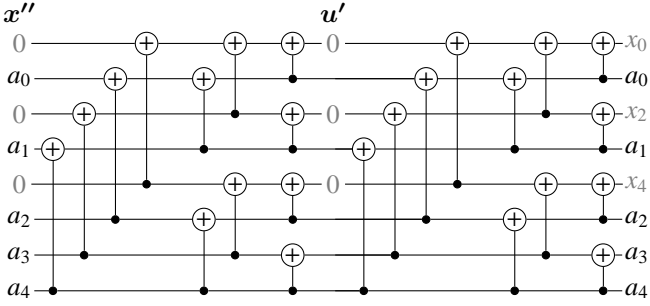
\begin{figure}
\centering
\newcommand{\ubit}[1]{$u_{#1}$}
\newcommand{\fbit}[1]{\color{gray}$u_{#1}$}
\newcommand{\ucw}[1]{$x_{#1}$}
\newcommand{\fcw}[1]{\color{gray}$x_{#1}$}
\newcommand{\ub}[1]{$#1$}
\newcommand{\fb}[1]{\color{gray}$#1$}

\begin{tikzpicture}[baseline=(n7s0.center)]

\usetikzlibrary{shapes,positioning,arrows,decorations.markings,fit}

\definecolor{varnode_fill}{RGB}{0,0,0}
\definecolor{chknode_fill}{RGB}{255,255,255}

\tikzset{
  chknode/.style={draw,fill=chknode_fill,circle,minimum size=0.3cm, inner sep=0},
  varnode/.style={draw,fill=varnode_fill,circle,minimum size=0.1cm, inner sep=0},
  sep/.style={rectangle,minimum width=0.25cm, inner sep=0},
  empty/.style={rectangle, inner sep=0},
  bit/.style={circle, inner sep = 0}
}

\matrix[row sep=1mm, column sep=1mm] {
  \node[bit] (n0s0) {\fb{0}}; & \node[empty] {}; & \node[empty] {}; &  \node[empty] {}; & \node[chknode] (n0s1) {$+$}; & \node[sep] (s10) {}; & \node[empty] {}; &	\node[chknode] (n0s2) {$+$}; &	\node[sep] (s20) {}; & \node[chknode] (n0s3) {$+$}; &  & \node[bit] (xn0s0) {\fb{0}}; & \node[empty] {}; & \node[empty] {}; &  \node[empty] {}; & \node[chknode] (xn0s1) {$+$}; & \node[sep] (s10) {}; & \node[empty] {}; &	\node[chknode] (xn0s2) {$+$}; &	\node[sep] (s20) {}; & \node[chknode] (xn0s3) {$+$}; & \node[bit] (xn0s4) {\fcw{0}}; \\
  \node[bit] (n4s0) {\ub{a_0}}; & \node[empty] {}; & \node[empty] {}; & \node[chknode] (n4s1) {$+$}; & \node[sep] (s14) {}; & \node[empty] {}; &   \node[chknode] (n4s2) {$+$}; &  \node[sep] (s24) {}; &  \node[empty] {}; & \node[varnode] (n4s3) {};	&  & \node[bit] (xn4s0) {}; & \node[empty] {}; & \node[empty] {}; & \node[chknode] (xn4s1) {$+$}; & \node[sep] (s14) {}; & \node[empty] {}; &   \node[chknode] (xn4s2) {$+$}; &  \node[sep] (s24) {}; &  \node[empty] {}; & \node[varnode] (xn4s3) {};	& \node[bit] (xn4s4) {\ub{a_0}}; \\
  \node[bit] (n2s0) {\fb{0}}; & & \node[chknode] (n2s1) {$+$}; & \node[sep] (s12) {}; && &&  \node[varnode] (n2s2) {};    &  \node[sep] (s22) {}; &   \node[chknode] (n2s3) {$+$}; & \node[bit] (n2s4) {}; &   \node[bit] (xn2s0) {\fb{0}}; &  & \node[chknode] (xn2s1) {$+$}; & \node[sep] (s12) {}; && &&  \node[varnode] (xn2s2) {};    &  \node[sep] (s22) {}; &   \node[chknode] (xn2s3) {$+$}; & \node[bit] (xn2s4) {\fcw{2}}; \\
  \node[bit] (n6s0) {\ub{a_1}}; & \node[chknode] (n6s1) {$+$}; & \node[sep] (s16) {}; & \node[empty] {};& \node[empty] {}; & \node[empty] {};&   \node[varnode] (n6s2) {};    &  \node[sep] (s26) {}; & \node[empty] {};&   \node[varnode] (n6s3) {};	&  &   \node[bit] (xn6s0) {}; & \node[chknode] (xn6s1) {$+$}; & \node[sep] (s16) {}; & \node[empty] {};& \node[empty] {}; & \node[empty] {};&   \node[varnode] (xn6s2) {};    &  \node[sep] (s26) {}; & \node[empty] {};&   \node[varnode] (xn6s3) {};	& \node[bit] (xn6s4) {\ub{a_1}}; \\
  \node[bit] (n1s0) {\fb{0}}; &&& & \node[varnode] (n1s1) {};    & \node[sep] (s11) {}; & &  \node[chknode] (n1s2) {$+$}; & \node[sep] (s21) {}; &  \node[chknode] (n1s3) {$+$}; &  &   \node[bit] (xn1s0) {\fb{0}}; &&& & \node[varnode] (xn1s1) {};    & \node[sep] (s11) {}; & &  \node[chknode] (xn1s2) {$+$}; & \node[sep] (s21) {}; &  \node[chknode] (xn1s3) {$+$}; &  \node[bit] (xn1s4) {\fcw{4}}; \\
  \node[bit] (n5s0) {\ub{a_2}}; && & \node[varnode] (n5s1) {};     & \node[sep] (s15) {}; & &  \node[chknode] (n5s2) {$+$}; & \node[sep] (s25) {}; &&  \node[varnode] (n5s3) {};	&  &   \node[bit] (xn5s0) {}; && & \node[varnode] (xn5s1) {};     & \node[sep] (s15) {}; & &  \node[chknode] (xn5s2) {$+$}; & \node[sep] (s25) {}; &&  \node[varnode] (xn5s3) {};	& \node[bit] (xn5s4) {\ub{a_2}}; \\
  \node[bit] (n3s0) {\ub{a_3}}; & & \node[varnode] (n3s1) {};    & \node[sep] (s13) {}; &&& &  \node[varnode] (n3s2) {};    & \node[sep] (s23) {}; &     \node[chknode] (n3s3) {$+$}; & &   \node[bit] (xn3s0) {}; & & \node[varnode] (xn3s1) {};    & \node[sep] (s13) {}; &&& &  \node[varnode] (xn3s2) {};    & \node[sep] (s23) {}; &     \node[chknode] (xn3s3) {$+$}; & \node[bit] (xn3s4) {\ub{a_3}}; \\
  \node[bit] (n7s0) {\ub{a_4}}; & \node[varnode] (n7s1) {};    & \node[sep] (s17) {}; &&& &  \node[varnode] (n7s2) {};    && \node[sep] (s27) {}; &    \node[varnode] (n7s3) {};	& &   \node[bit] (xn7s0) {}; & \node[varnode] (xn7s1) {};    & \node[sep] (s17) {}; &&& &  \node[varnode] (xn7s2) {};    && \node[sep] (s27) {}; &    \node[varnode] (xn7s3) {};	& \node[bit] (xn7s4) {\ub{a_4}}; \\
};
\path[-] (n0s0) edge (n0s1) (n0s1) edge (n0s2) (n0s2) edge (n0s3) (n0s3) edge (xn0s0);
\path[-] (n1s0) edge (n1s1) (n1s1) edge (n1s2) (n1s2) edge (n1s3) (n1s3) edge (xn1s0);
\path[-] (n2s0) edge (n2s1) (n2s1) edge (n2s2) (n2s2) edge (n2s3) (n2s3) edge (xn2s0);
\path[-] (n3s0) edge (n3s1) (n3s1) edge (n3s2) (n3s2) edge (n3s3) (n3s3) edge (xn3s1);
\path[-] (n4s0) edge (n4s1) (n4s1) edge (n4s2) (n4s2) edge (n4s3) (n4s3) edge (xn4s1);
\path[-] (n5s0) edge (n5s1) (n5s1) edge (n5s2) (n5s2) edge (n5s3) (n5s3) edge (xn5s1);
\path[-] (n6s0) edge (n6s1) (n6s1) edge (n6s2) (n6s2) edge (n6s3) (n6s3) edge (xn6s1);
\path[-] (n7s0) edge (n7s1) (n7s1) edge (n7s2) (n7s2) edge (n7s3) (n7s3) edge (xn7s1);

\path[-] (n0s1) edge (n1s1);
\path[-] (n2s1) edge (n3s1);
\path[-] (n4s1) edge (n5s1);
\path[-] (n6s1) edge (n7s1);

\path[-] (n0s2) edge (n2s2);
\path[-] (n1s2) edge (n3s2);
\path[-] (n4s2) edge (n6s2);
\path[-] (n5s2) edge (n7s2);

\path[-] (n0s3) edge (n4s3);
\path[-] (n1s3) edge (n5s3);
\path[-] (n2s3) edge (n6s3);
\path[-] (n3s3) edge (n7s3);

%%%%%%%%%%%%%%%%%%%%%%%%%%%%%%%%%%%%%%%%%

\path[-] (xn0s0) edge (xn0s1) (xn0s1) edge (xn0s2) (xn0s2) edge (xn0s3) (xn0s3) edge (xn0s4);
\path[-] (xn1s0) edge (xn1s1) (xn1s1) edge (xn1s2) (xn1s2) edge (xn1s3) (xn1s3) edge (xn1s4);
\path[-] (xn2s0) edge (xn2s1) (xn2s1) edge (xn2s2) (xn2s2) edge (xn2s3) (xn2s3) edge (xn2s4);
\path[-] (xn3s0) edge (xn3s1) (xn3s1) edge (xn3s2) (xn3s2) edge (xn3s3) (xn3s3) edge (xn3s4);
\path[-] (xn4s0) edge (xn4s1) (xn4s1) edge (xn4s2) (xn4s2) edge (xn4s3) (xn4s3) edge (xn4s4);
\path[-] (xn5s0) edge (xn5s1) (xn5s1) edge (xn5s2) (xn5s2) edge (xn5s3) (xn5s3) edge (xn5s4);
\path[-] (xn6s0) edge (xn6s1) (xn6s1) edge (xn6s2) (xn6s2) edge (xn6s3) (xn6s3) edge (xn6s4);
\path[-] (xn7s0) edge (xn7s1) (xn7s1) edge (xn7s2) (xn7s2) edge (xn7s3) (xn7s3) edge (xn7s4);

\path[-] (xn0s1) edge (xn1s1);
\path[-] (xn2s1) edge (xn3s1);
\path[-] (xn4s1) edge (xn5s1);
\path[-] (xn6s1) edge (xn7s1);

\path[-] (xn0s2) edge (xn2s2);
\path[-] (xn1s2) edge (xn3s2);
\path[-] (xn4s2) edge (xn6s2);
\path[-] (xn5s2) edge (xn7s2);

\path[-] (xn0s3) edge (xn4s3);
\path[-] (xn1s3) edge (xn5s3);
\path[-] (xn2s3) edge (xn6s3);
\path[-] (xn3s3) edge (xn7s3);

\node[above=0cm of n0s0] {\mvec{$x''$}};
\node[above=0cm of xn0s0] {\mvec{$u'$}};

\end{tikzpicture}
\caption{Systematic encoding with bit-reversal.}
\label{fig:ssc:bit-rev}
\end{figure}

To illustrate the encoding method, Fig.~\ref{fig:ssc:bit-rev} shows the encoding process for an (8, 5) polar code with bit-reversal. ($x''_0$, $x''_2$, $x''_4$) are frozen and set to 0 according to the bit-reversed indices of the least reliable bits; and ($x''_1$, $x''_3$, $x''_5$, $x''_6$, $x''_7$) are set to the information bits ($a_0$, $a_1$, $a_2$, $a_3$, $a_4$). $\mvec{x''}$ is encoded using $G$ to obtain the vector $\mvec{u'}$, in which the bits ($u'_0$, $u'_2$, $u'_4$) are then set to zero. The resulting $\mvec{u'}$ is encoded again yielding the systematic codeword $\mvec{x}$, which is transmitted over the channel sequentially, i.e. $x_0$ then $x_1$ and so on. An encoder that does not use bit-reversal will function in the same manner, except that the frozen bit indices will be (0, 1, 2).
An SSC decoder with $P = 2$ will output ($\estx{0}$, $\estx{1}$, $\estx{2}$, $\estx{3}$) then ($\estx{4}$, $\estx{5}$, $\estx{6}$, $\estx{7}$), i.e. the output of the decoder is ($\estx{0}$, $\esta{0}$, $\estx{2}$, $\esta{1}$) then ($\estx{4}$, $\esta{2}$, $\esta{3}$, $\esta{4}$) where the source data estimate appears in the correct order.

\section{Proposed Algorithm}
\label{sec:algorithm}
In this section we explore more constituent codes that can be decoded directly and present the associated specialized decoding algorithms. We present three new corresponding node types:  a single-parity-check-code node, a repetition-code node, and a special node whose left child corresponds to a repetition code and its right to a single-parity-check code. We also present node mergers that reduce decoder latency and summarize all the functions the new decoder must  be able to perform. Finally, we study the effect of quantization on the error-correction performance of the proposed algorithm.

It should be noted that all the transformations and mergers presented in this work preserve the polar code, i.e. they do not alter the locations of frozen and information bits. While some throughput improvement is possible via some code modifications, the resulting polar code diverges from the optimal one constructed according to \cite{Tal2011a}.

To keep the results in this section practical, we use $P$ as a resource constraint parameter, similar to \cite{Leroux2013}. However, since new node types are introduced, the notion of a processing element (PE) might not apply in certain cases. Therefore, we redefine $P$ so that $2P$ is the maximum number of memory elements that can be accessed simultaneously. Since each PE has two inputs, $P$ PEs require $2P$ input values and the two definitions for $P$ are compatible. In addition, $P$ is as a power of two as in \cite{Leroux2013}.

\subsection{Single-Parity-Check Nodes $\nspc$}
\label{sec:algorithm:spc}
In any polar code of rate $(N-1)/N$, the frozen bit is always $u_0$ rendering the code a single-parity check (SPC) code, which can be observed in Fig.~\ref{fig:pc4}. While the dimension of an SPC code is $N-1$, for which exhaustive-search ML decoding is impractical; optimal ML decoding can still be performed with very low complexity \cite{Snyders1989}:  the hard-decision estimate and the parity of the input are calculated; then the estimate of the least reliable bit is flipped if the parity constraint is not satisfied.
The hard-decision estimate of the soft-input values is calculated using
\[
\text{HD}[i] = \begin{cases}
0, & \text{when } \mvec{\alpha}_v \geq 0;\\
1, & \text{otherwise.}
\end{cases}
\]
The parity of the input is calculated as
\begin{equation}
\label{eq:spc:parity}
\text{parity} = \bigoplus_{i = 0}^{N_v-1} \text{HD}[i].
\end{equation}
The index of the least reliable input is found using
\[
j = \argmin_i |\mvec{\alpha}_v[i]|.
\]
Finally, the output of the node is
\begin{equation}
\label{eq:spc:beta}
\mvec{\beta}_v[i] = \begin{cases}
\text{HD}[i] \oplus \text{parity}, & \text{when } i = j;\\
\text{HD}[i], & \text{otherwise.}
\end{cases}
\end{equation}

The resulting node can decode an SPC code of length $N_v > 2P$ in $N_v/(2P) + c$ steps, where $c \geq 1$ since at least one step is required to correct the least reliable estimate and others might be used for pipelining; whereas an SSC decoder requires $2 \sum_{i = 1}^{\log_2 N_v} \lceil 2^i / (2P) \rceil$ steps. For example, for an SPC constituent code of length 4096, $P = 256$, and $c = 4$, the specialized SPC decoder requires 12 steps, whereas the SSC decoder requires 46 steps.
For constituent codes of length $\leq 2P$ the decoder can provide an output immediately, or after a constant number of time steps if pipelining is used.

Large SPC constituent codes are prevalent in high-rate polar codes and a significant reduction in latency can be achieved if they are decoded quickly. Table~\ref{tab:spc-dist} lists the number of SPC nodes, binned by size, in three different polar codes: (32768, 29492), (32768, 27568), and a lower-rate (32768, 16384), all constructed for an AWGN channel with a noise variance $\sigma^2=0.1936$.
Comparing the results for the three codes, we observed that the total number of nodes decreases as the rate increases. The distribution of SPC nodes by length is also affected by code rate: the proportion of large SPC nodes decreases as the rate decreases.

\begin{table}[t]
\centering
\caption{Number of all nodes and of SPC nodes of different sizes in three polar codes of length 32768 and rates 0.9, 0.8413, and 0.5.}
\begin{tabular}{cccccc}
  \toprule
  \multirow{2}{*}{Code} & \multirow{2}{*}{All} & \multicolumn{4}{c}{SPC, $N_v \in$}\\
  \cmidrule{3-6}
   & & $(0, 8]$ & $(8, 64]$ & $(64, 256]$ & $(256, 32768]$\\
  \midrule
  (32768, 29492) & 2065 & 383 & 91 & 17 & 13\\
  (32768, 27568) & 3421 & 759 & 190 & 43 & 10\\
  (32768, 16384) & 9593 & 2240 & 274 & 19 & 1\\
  \bottomrule
\end{tabular}
\label{tab:spc-dist}
\end{table}

A generalized version of the single-parity-check nodes, called caterpillar nodes, was presented in \cite{Alamdar-Yazdi2012} and was shown to improve throughput of SSC by 11--14\% when decoding polar codes transmitted over the binary erasure channel (BEC) without resource constraints.

\subsection{Repetition Nodes $\nrep$}
\label{sec:algorithm:rep}

Another type of constituent codes that can be decoded more efficiently than using tree traversal is repetition codes, in which only the last bit is not frozen. The decoding algorithm starts by summing all input values. Threshold detection is performed via sign detection, and the result is replicated and used as the constituent decoder's final output:
\begin{equation}
\label{eq:rep}
\mvec{\beta}_v[i] = \begin{cases}
0, & \text{when } \sum_j \mvec{\alpha}_v[j] \geq 0;\\
1, & \text{otherwise.}
\end{cases}
\end{equation}

\begin{table}[t]
\centering
\caption{Number of all nodes and of repetition nodes of different sizes in three polar codes of length 32768 and rates 0.9, 0.8413, and 0.5.}
\begin{tabular}{ccccc}
  \toprule
  \multirow{2}{*}{Code} & \multirow{2}{*}{All} & \multicolumn{3}{c}{Repetition, $N_v \in$}\\
  \cmidrule{3-5}
   & & $(0, 8]$ & $(8, 16]$ & $(16, 32768]$\\
  \midrule
  (32768, 29492) & 3111 & 474 & 30 & 0 \\
  (32768, 27568) & 5501 & 949 & 53 & 0 \\
  (32768, 16384) & 10381 & 2290 & 244 & 0 \\
  \bottomrule
\end{tabular}
\label{tab:rep-dist}
\end{table}

The decoding method \eqref{eq:rep} requires $N_v/(2P)$ steps to calculate the sum and $N_v/(2P)$ steps to set the output, in addition to any extra steps required for pipelining. Two other methods employing prediction can be used to decrease latency. The first sets all output bits to 0 while accumulating the inputs, and writes the output again only if the sign of the sum is negative. The average latency of this method is 75\% that of \eqref{eq:rep}. The second method sets half the output words to all 0 and the other half to all 1, and corrects the appropriate words when the sum is known. The resulting latency is 75\% that of \eqref{eq:rep}. However, since the high-rate codes of interest do not have any large repetition constituent codes, we chose to use \eqref{eq:rep} directly.

Unlike SPC constituent codes, repetition codes are more prevalent in lower-rate polar codes as shown in Table~\ref{tab:rep-dist}. Moreover, for high-rate codes, SPC nodes have a more pronounced impact on latency reduction. This can be observed in tables \ref{tab:spc-dist} and \ref{tab:rep-dist}, which show that the total number of nodes in the decoder tree is significantly smaller when only SPC nodes are introduced than when only repetition nodes are introduced, indicating a smaller tree and lower latency. Yet, the impact of repetition nodes on latency is measurable; therefore, we use them in the decoder.

\subsection{Repetition-SPC Nodes $\nrepspc$}
\label{sec:algorithm:rep-spc}
When enumerating constituent codes with $N_v \leq 8$ and $0 < k_v < 8$ for the (32768, 27568) and (32768, 29492) codes, three codes dominated the listing: the SPC code, the repetition code, and a special code whose left constituent code is a repetition code and its right an SPC one, denoted $\nrepspc$. The other constituent codes accounted for 6\% and 12\% in the two polar codes, respectively. Since $\nrepspc$ codes account for 28\% and 25\% of the total $\nr$ nodes of length 8 in the two aforementioned codes, efficiently decoding them would have a significant impact on latency. This can be achieved by using two SPC decoders of length $4$, $\text{SPC}_0$ and $\text{SPC}_1$, whose inputs are calculated assuming the output of the repetition code is 0 and 1, respectively. Simultaneously, the repetition code is decoded and its output is used to generate the $\nrepspc$ output using either the output of $\text{SPC}_0$ or $\text{SPC}_1$ as appropriate.

While this code can be decoded using an exhaustive-search ML decoder, the proposed decoder has a significantly lower complexity.

\subsection{Node Mergers}
\label{sec:algorithm:mergers}
The $\nrepspc$ node merges an $\nrep$ and an $\nspc$ node to reduce latency. Similarly, it was mentioned in \cite{Sarkis2013} that $\nr$ nodes need not calculate the input to a child node if it is an $\nz$ node. Instead, the input to the right child is directly calculated.

Another opportunity for a node merger arises when a node's right child directly provides $\mvec{\beta}_r$ without tree traversal: the calculation of $\mvec{\alpha}_r$, $\mvec{\beta}_r$, and $\mvec{\beta}_v$ can all be performed in one step, halving the latency. This is also applicable for nodes where $N_v > 2P$: $P$ values of $\mvec{\alpha}_r$ are calculated and used to calculate $P$ values of $\mvec{\beta}_r$, which are then used to calculate $2P$ values of $\mvec{\beta}_v$ until all values have been calculated.

This can be expanded further when the left node is $\nz$. Since $\mvec{\beta}_l$ is known a priori to be a zero vector, $\mvec{\alpha}_r$ can be immediately calculated once $\mvec{\alpha}_v$ is available and $\mvec{\beta}_r$ is combined with the zero vector to obtain $\mvec{\beta}_v$.

In all the codes that were studied, $\nr$, $\no$, and $\nspc$ were the only nodes to be observed as right children; and $\no$ and $\nspc$ are the only two that can be merged with their parent.

\subsection{Required Decoder Functions}
\label{sec:algorithm:functions}
\begin{table}[t]
\centering
\caption{A listing of the different functions performed by the proposed decoder.}
\begin{tabular}{ll}
  \toprule
  Name & Description\\
  \midrule
  F & calculate $\mvec{\alpha}_l$ \eqref{eq:sc:ms:f}.\\
  G & calculate $\mvec{\alpha}_r$ \eqref{eq:sc:ms:g}.\\
  COMBINE & combine $\mvec{\beta}_l$ and $\mvec{\beta}_r$ \eqref{eq:xor}.\\
  COMBINE-0R & same as COMBINE, but with $\mvec{\beta}_l = 0$.\\
  G-0R & same as G, but assuming $\mvec{\beta}_l = 0$.\\
  P-R1 & calculate $\mvec{\beta}_v$ using \eqref{eq:sc:ms:g}, \eqref{eq:threshold}, then \eqref{eq:xor}.\\
  P-RSPC & calculate $\mvec{\beta}_v$ using \eqref{eq:sc:ms:g}, \eqref{eq:spc:beta}, then \eqref{eq:xor}.\\
  P-01 & same as P-R1, but assuming $\mvec{\beta}_l = 0$.\\
  P-0SPC & same as P-RSPC, but assuming $\mvec{\beta}_l = 0$.\\
  ML & calculate $\mvec{\beta}_v$ using exhaustive-search ML decoding.\\
  REP & calculate $\mvec{\beta}_v$ using \eqref{eq:rep}.\\
  REP-SPC & calculate $\mvec{\beta}_v$ as in Section~\ref{sec:algorithm:rep-spc}.\\
  \bottomrule
\end{tabular}
\label{tab:functions}
\end{table}

As a result of the many types of nodes and the different mergers, the decoder must perform many functions. Table~\ref{tab:functions} lists these 12 functions. For notation, 0, 1, and R are used to denote children with constituent code rates of 0, 1, and $R$, respectively. Having a left child of rate 0 allows the calculation of $\mvec{\alpha}_r$ directly from $\mvec{\alpha}_v$ as explained earlier. It is important to make this distinction since the all-zero output of a rate 0 code is not stored in the decoder memory. In addition, having a right child of rate 1 allows the calculation of $\mvec{\beta}_v$ directly once $\mvec{\beta}_l$ is known. A P- prefix indicates that the message to the parent, $\mvec{\beta}_v$, is calculated without explicitly visiting the right child node.

We note the absence of $\nz$ and $\no$ node functions: the former due to directly calculating $\mvec{\alpha}_r$ and the latter to directly calculating $\mvec{\beta}_v$ from $\mvec{\alpha}_r$.

\subsection{Performance with Quantization}
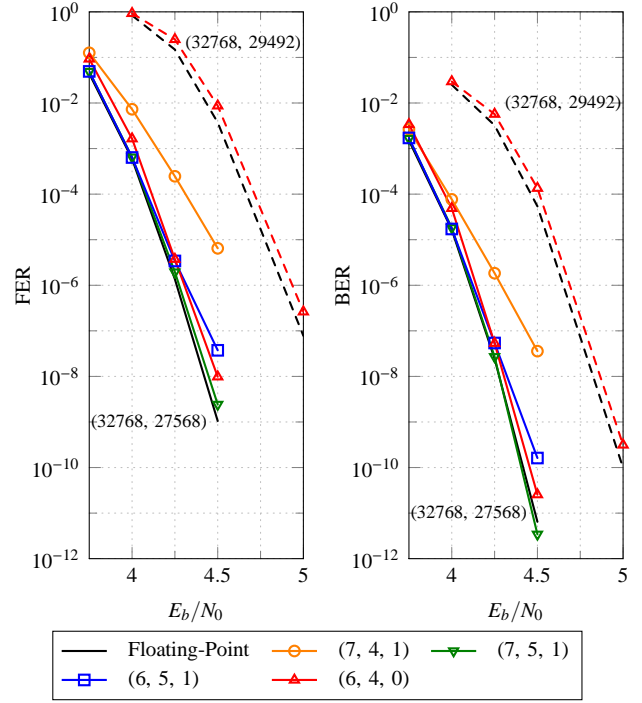
\begin{figure}[t]
\centering
\centering
\usetikzlibrary{plotmarks}

\definecolor{darkgreen}{RGB}{0, 128, 0}
\definecolor{mypurple}{RGB}{153, 71, 155}

\begin{tikzpicture}

 \begin{semilogyaxis}
    [
    width=0.5\columnwidth,
    height=\columnwidth,
    line cap=round,
    every axis/.append style={font=\footnotesize},
    every axis y label/.style={at={(ticklabel cs:0.5)},rotate=90},
    every axis x label/.style={at={(ticklabel cs:0.5)},anchor=near ticklabel},
    every axis plot/.append style={thick},
    xmin=3.75,
    xmax=5,
    xlabel={$E_b/N_0$},
    xtick={4, 4.5, 5},
    minor x tick num={1},
    ymax=1.0,
    ymin=1.0e-12,
    ytickten={0,-2,...,-12},
    minor ytick={1e-1, 1e-3, 1e-5, 1e-7, 1e-9, 1e-11},
    ylabel={FER},
    legend columns = 3,
    legend style={
      anchor={center},
      cells={anchor=west},
      column sep= 2mm,
      font=\footnotesize,
    },
    legend to name=quant-legend,
    grid=both,
    grid style=dotted,
    ]

    \addplot[color=black] table[x=snr_db,y=FER] {data/pc-ssc32k.o7727};
    \addlegendentry{Floating-Point}

    \addplot[color=orange,mark=o] table[x=snr_db,y=FER] {data/mlssc32k.o8127};
    \addlegendentry{(7, 4, 1)}

    \addplot[color=darkgreen,mark=triangle, mark options={rotate=60}] table[x=snr_db,y=FER] {data/mlssc32k.o8122};
    \addlegendentry{(7, 5, 1)}

    \addplot[color=blue, mark=square] table[x=snr_db, y=FER] {data/mlssc32k.o8124};
    \addlegendentry{(6, 5, 1)}

    \addplot[color=red, mark=triangle] table[x=snr_db, y=FER] {data/mlssc32k.o8131};
    \addlegendentry{(6, 4, 0)}

    \addplot[color=black, dashed] table[x=snr_db, y=FER] {data/mlssc32k.o8134};
    \addplot[color=red, dashed, mark=triangle, mark options={solid}] table[x=snr_db, y=FER] {data/mlssc32k.o8135};

    \node[] at (axis cs:4.1,1e-09) {\scriptsize(32768, 27568)};
    \node[] at (axis cs:4.65,0.2) {\scriptsize(32768, 29492)};

  \end{semilogyaxis}
\end{tikzpicture}
\begin{tikzpicture}

 \begin{semilogyaxis}
    [
    width=0.5\columnwidth,
    height=\columnwidth,
    line cap=round,
    every axis/.append style={font=\footnotesize},
    every axis y label/.style={at={(ticklabel cs:0.5)},rotate=90},
    every axis x label/.style={at={(ticklabel cs:0.5)},anchor=near ticklabel},
    every axis plot/.append style={thick},
    xmin=3.75,
    xmax=5,
    xlabel={$E_b/N_0$},
    xtick={4, 4.5, 5},
    minor x tick num={1},
    ymax=1.0,
    ymin=1.0e-12,
    ytickten={0,-2,...,-12},
    minor ytick={1e-1, 1e-3, 1e-5, 1e-7, 1e-9, 1e-11},
    ylabel={BER},
    grid=both,
    grid style=dotted,
    ]

    \addplot[color=black] table[x=snr_db,y=BER] {data/pc-ssc32k.o7727};

    \addplot[color=orange,mark=o] table[x=snr_db,y=BER] {data/mlssc32k.o8127};

    \addplot[color=darkgreen,mark=triangle, mark options={rotate=60}] table[x=snr_db,y=BER] {data/mlssc32k.o8122};

    \addplot[color=blue, mark=square] table[x=snr_db, y=BER] {data/mlssc32k.o8124};

    \addplot[color=red, mark=triangle] table[x=snr_db, y=BER] {data/mlssc32k.o8131};

    \addplot[color=black, dashed] table[x=snr_db, y=BER] {data/mlssc32k.o8134};
    \addplot[color=red, dashed, mark=triangle, mark options={solid}] table[x=snr_db, y=BER] {data/mlssc32k.o8135};

    \node[] at (axis cs:4.1,1e-11) {\scriptsize(32768, 27568)};
    \node[] at (axis cs:4.65,0.01) {\scriptsize(32768, 29492)};

  \end{semilogyaxis}
\end{tikzpicture}

\ref{quant-legend}
\caption{Effect of quantization on the error-correction performance of the (32768, 27568) and (32768, 29492) codes.}
\label{fig:quant-perf-84}
\end{figure}

Fig.~\ref{fig:quant-perf-84} shows the effect of quantization on the (32768, 27568) polar code that was constructed for $E_b/N_0 = 4.5$ dB. The quantization numbers are presented in $(W, W_C, F)$ format, where $W$ is total number of quantization bits for internal LLRs, $W_c$ for channel LLRs, and $F$ is the number of fractional bits. Since the proposed algorithm does not perform any operations that increase the number of fractional bits---only the integer ones---we use the same number of fractional bits for both internal and channel LLRs.

From the figure, it can be observed that using a $(7, 5, 1)$ quantization scheme yields performance extremely close to that of the floating-point decoder.
Decreasing the range of the channel values to three bits by using the $(7, 4, 1)$ scheme significantly degrades performance.
While completely removing fractional bits, $(6, 4, 0)$, yields performance that remains within 0.1 dB of the floating-point decoder throughout the entire $E_b/N_0$ range. This indicates that the decoder needs four bits of range for the channel LLRs.
Keeping the channel LLR quantization the same, but reducing the range of the internal LLRs by one bit and using $(6, 5, 1)$ quantization does not affect the error-correction performance for $E_b/N_0 < 4.25$. After that point however, the performance starts to diverge from that of the floating-point decoder. Therefore, the range of internal LLR values increases in importance as $E_b/N_0$ increases. Similarly, using $(6, 4, 0)$ quantization proved sufficient for decoding the (32768, 29492) code.

From these results, we conclude that minimum number of integer quantization bits required is six for the internal LLRs and four for the channel ones and that fractional bits have a small effect on the performance of the studied polar codes. The $(6, 4, 0)$ scheme offers lower memory use for a small reduction in performance and would be the recommended scheme for a practical decoder for high-rate codes. For the rest of this work, we use both the $(6, 4, 0)$ and $(7, 5, 1)$ schemes to illustrate the performance-complexity trade off between them.

\label{sec:algorithm:performance}

\subsection{Latency Compared to ML-SSC Decoding}
The different nodes have varying effects on the latency.
Table~\ref{tab:latency} lists the latency, in clock cycles, of the ML-SSC
decoder without utilizing any of the new node types when decoding a (32768,
29492) code. It then lists the latency of that decoder with the addition of each of the different node types individually, and finally with all of the nodes. Since this is a high rate code, $\nrep$ nodes have a small effect on latency. An ML-SSC decoder with $\nrepspc$ nodes has 89.7\% the latency of the regular ML-SSC decoder, and one with $\nspc$ node has 63.6\% the latency. Finally, the proposed decoder with all nodes has 54\% the latency of the ML-SSC decoder. From these results, we conclude that $\nspc$ nodes have the largest effect on reducing the latency of decoding this code; however, other nodes also contribute measurably.

\begin{table}
  \centering
  \caption{Latency of ML-SSC decoding of the (32768, 29492) code and the effect of using additional nodes types on it.}
  \begin{tabular}{c c c c c}
    \toprule
    None & SPC & REP-SPC & REP & All\\
    \midrule
    5286 & 3360 & 4742 & 5042 & 2847\\
    \bottomrule
  \end{tabular}
  \label{tab:latency}
\end{table}

\section{Architecture: Top-Level}
\label{sec:architecture:top}
As mentioned earlier, Table~\ref{tab:functions} lists the 12 functions performed by the decoder. Deducing which function to perform online would require complicated controller logic. Therefore, the decoder is provided with an offline-calculated list of functions to perform. This does not reduce the decoder's flexibility as a new set of functions corresponding to a different code can be loaded at any time. To further simplify implementation, we present the decoder with a list of instructions, with each instruction composed of the function to be executed, and a value indicating whether the function is associated with a right or a left child in the decoder tree. An instruction requires 5 bits to store: 4 bits to encode the operation and 1 bit to indicate child association. For the $N = 32768$ codes in this work, the maximum instruction memory size was set to $3000 \times 5$ bits, which is smaller than the 32768 bits required to directly store a mask of the frozen-bit locations. This list of instructions can be viewed as a program executed by a specialized micro-processor, in this case, the decoder.

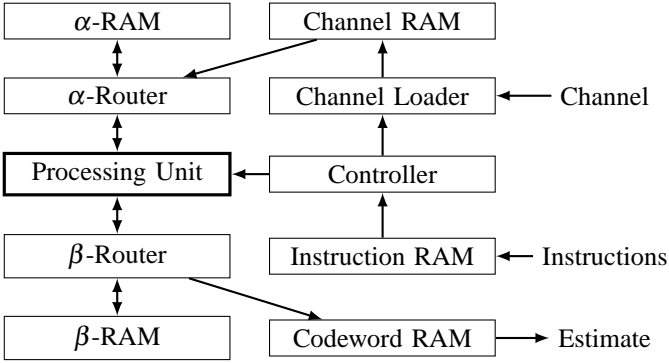
\begin{figure}
\centering
\begin{tikzpicture}

\usetikzlibrary{shapes,positioning,arrows,decorations.markings,fit}

\tikzset{
  block/.style={draw,rectangle, minimum width = 3cm},
  >={latex},
}

\matrix[row sep=5mm, column sep=5mm] {

\node[block] (alpha_ram) {$\alpha$-RAM}; & \node[block] (chan_ram) {Channel RAM}; & \\
\node[block] (alpha_router) {$\alpha$-Router}; & \node[block] (chan_loader) {Channel Loader}; &  \node[rectangle] (chan) {Channel};\\
\node[block,very thick] (processor) {Processing Unit}; & \node[block] (controller) {Controller}; & \\
\node[block] (beta_router) {$\beta$-Router}; & \node[block] (inst_ram) {Instruction RAM};& \node[rectangle] (instructions) {Instructions}; \\
\node[block] (beta_ram) {$\beta$-RAM}; & \node[block] (codeword_ram) {Codeword RAM}; & \node[rectangle] (estimate) {Estimate};\\
};

\draw[thick,->] (chan) -- (chan_loader);

\draw[thick,->] (instructions) -- (inst_ram);
\draw[thick,->] (inst_ram) -- (controller);
\draw[thick,->] (controller) -- (processor);
\draw[thick,->] (controller) -- (chan_loader);
\draw[thick,->] (chan_loader) -- (chan_ram);

\draw[thick,<->] (alpha_router) -- (alpha_ram);
\draw[thick,->] (chan_ram) -- (alpha_router);
\draw[thick,<->] (alpha_router) -- (processor);
\draw[thick,<->] (beta_router) -- (processor);
\draw[thick,<->] (beta_router) -- (beta_ram);
\draw[thick,->] (beta_router) -- (codeword_ram);

\draw[thick,->] (codeword_ram) -- (estimate);

\end{tikzpicture}
\caption{Top-level architecture of the decoder.}
\label{fig:arch}
\end{figure}

With such a view, we present the overall architecture of our decoder, shown in Fig.~\ref{fig:arch}.
At the beginning, the instructions (program) are loaded into the instruction RAM (instruction memory) and fetched by the controller (instruction decoder). The controller then signals the channel loader to load channel LLRs into memory, and data processing unit (ALU) to perform the correct function. The processing unit accesses data in $\mvec{\alpha}-$ and $\mvec{\beta}-$ RAMs (data memory). The estimated codeword is buffered into the codeword RAM which is accessible from outside the decoder.

By using a pre-compiled list of instructions, the controller is reduced to fetching and decoding instructions, tracking which stage is currently decoded, initiating channel LLR loading, and triggering the processing unit.

Before discussing the details of the decoder architecture, it should be noted that this work presents a complete decoder, including all input and output buffers needed to be flexible. While it is possible to reduce the size of the buffers, this is accompanied by a reduction in flexibility and limits the range of codes which can be decoded at full throughput, especially at high code rates. This trade off is explored in more detail in sections \ref{sec:data:channel-ram} and \ref{sec:implementation}.

\section{Architecture: Data Loading and Routing}
\label{sec:architecture:data}
When designing the decoder, we have elected to include the required input and output buffers in addition to the buffers required to store internal results. To enable data loading while decoding and achieve the maximum throughput supported by the algorithm, $\alpha$ values were divided between two memories: one for channel $\alpha$ values and the other for internal ones as described in sections \ref{sec:data:channel-ram} and \ref{sec:data:alpha-ram}, respectively. Similarly, $\beta$ values were divided between two memories as discussed in sections \ref{sec:data:beta-ram} and \ref{sec:data:codeword-ram}. Finally, routing of data to and from the processing unit is examined in Section~\ref{sec:data:routing}.

Since high throughput is the target of this design, we choose to improve timing and reduce routing complexity at the expense of logic and memory use.

\subsection{Channel $\alpha$ Values}
\label{sec:data:channel-ram}
Due to the lengths of polar codes with good error-correction performance, it is not practical to present all the channel output values to the decoder simultaneously. For the proposed design, we have settled to provide the channel output in groups of 32 LLRs; so that for a code of length 32768, 1024 clock cycles are required to load one frame in the channel RAM. Since the codes of rates 0.8413 and 0.9 require 3631 and 2847 clock cycles to decode, respectively, stalling the decoder while a new frame is loaded will reduce throughput by more than 25\%. Therefore, loading a new frame while currently decoding another is required to prevent throughput loss.

The method employed in this work for loading a new frame while decoding is to use a dual-port RAM that provides enough memory to store two frames. The write port of the memory is used by the channel loader to write the new frame; while the read port is used by the $\alpha$-router to read the current frame. Once decoding of the current frame is finished, the reading and writing locations in the channel RAM are swapped and loading of the new frame begins. This method was selected as it allowed full throughput decoding of both rate 0.8413 and 0.9 codes without the need for a faster second write clock while maintaining a reasonable decoder input bus width of $32 \times 5 = 160$ bits, where five quantization bits are used for the channel values, or 128 bits when using (6, 4, 0) quantization. Additionally, channel data can be written to the decoder at a constant rate by utilizing handshaking signals.

The decoder operates on $2P$ channel $\alpha$-values simultaneously, requiring access to a $2 * 256 * 5 = $ 2560-bit read bus. In order for the channel RAM to accommodate such a requirement while keeping the input bus width within practical limits, it must provide differently sized read and write buses. One approach is to use a very wide RAM and utilize a write mask; however, such wide memories are discouraged from an implementation perspective. Instead, multiple RAM banks, each has the same width as that of the input bus, are used. Data is written to one bank at a time, but read from all simultaneously. The proposed decoder utilizes $2 * 256 / 32 = 16$ banks each with a depth of 128 and a width of $32*5 = 160$ bits.

This memory cannot be merged with the one for the internal $\alpha$ values without stalling the decoder to load the new frame as the latter's two ports can be used by the decoder simultaneously and will not be available for another write operation.

Another method for loading-while-decoding is to replace the channel values once they are no longer required. This occurs after 2515 and 2119 clock cycles, permitting the decoder 1116 and 728 clock cycles in which to load the new frame for the $R = 0.8413$ and $R = 0.9$ codes, respectively. Given these timing constraints, the decoder is provided sufficient time to decode the rate 0.8413 code, but not the rate 0.9 one, at full throughput. To decode the latter, either the input bus width must be increased, which might not be possible given design constraints, or a second clock, operating faster than the decoder's, must be utilized for the loading operation. This approach sacrifices the flexibility of decoding very high-rate codes for a reduction in the channel RAM size. The impact of this compromise on implementation complexity is discussed in Section~\ref{sec:implementation}.

\subsection{Internal $\alpha$ Values}
\label{sec:data:alpha-ram}
The $f$ \eqref{eq:sc:ms:f} and $g$ \eqref{eq:sc:ms:g} functions are the only two components of the decoder that generate $\alpha$ values as output: each function accepts two $\alpha$ values as inputs and produces one. Since up to $P$ such functions are employed simultaneously, the decoder must be capable of providing $2P$ $\alpha$ values and of writing $P$ values. To support such a requirement, the internal $\alpha$ value RAM, denoted $\alpha$-RAM, is composed of two $P$-LLR wide memories. A read operation provides data from both memories; while a write operation only updates one.
Smaller decoder stages, which require fewer than $2P$ $\alpha$ values, are still assigned a complete memory word in each memory. This is performed to reduce routing and multiplexing complexity as demonstrated in \cite{Leroux2013}. Each memory can be composed of multiple RAM banks as supported by the implementation technology.

Since read from and write to $\alpha$-RAM operations can be performed simultaneously, it is possible to request a read operation from the same location that is being written. In this case, the memory must provide the most recent data. To provide this functionality for synchronous RAM, a register is used to buffer newly written data and to provide it when the read and write addresses are the same \cite{Leroux2013}.

\subsection{Internal $\beta$ Values}
\label{sec:data:beta-ram}
The memory used to store internal $\beta$ values needs to offer greater flexibility than $\alpha$-RAM, as some functions, such as COMBINE, generate $2P$ bits of $\beta$ values while others, such as ML and REP, generate $P$ or fewer bits.

The $\beta$-RAM is organized as two dual-port memories that are $2P$ bits wide each. One memory stores the output of left children while the other that of right ones. When a read operation is requested, data from both memories is read and either the lower or the upper half from each memories is selected according to whether the read address is even or odd. Similar to the $\alpha$ memories, the $\beta$ memories can be composed of multiple banks each.

Since $\beta$-RAM is read from and written to simultaneously, using the second port of a narrower dual-port RAM and writing to two consecutive addresses to improve memory utilization is not possible as it would interfere with the read operation and reduce throughput.

\subsection{Estimated Codeword}
\label{sec:data:codeword-ram}
The estimated codeword is generated $2P = 512$ bits at a time. These estimated bits are stored in the codeword RAM in order to enable the decoder to use a bus narrower than 512 bits to convey its estimate and to start decoding the following frame immediately after finishing the current. In addition, buffering the output allows the estimate to be read at a constant rate.
The codeword RAM is a simple dual-port RAM with a $2P = 512$-bit write bus and a 256-bit read bus and is organized as $N/(2P) = 64$ words of 512 bits.

Similar to the case of $\alpha$ value storage, this memory must remain separate from the internal $\beta$ memory in order to support decoding at full speed; otherwise, decoding must be stalled while the estimated codeword is read due to lack of available ports in RAM.

\subsection{Routing}
\label{sec:data:routing}
Since both $\alpha$ and $\beta$ values are divided between two memories, some logic is required to determine which memory to access, which is provided by the $\alpha$- and $\beta$- routers.

The $\alpha$-router receives stage and word indices, determines whether to fetch data from the channel or $\alpha$-RAM, and calculates the read address. Only $\alpha$-RAM is accessible for write operations through the $\alpha$-router . Similarly, the $\beta$-router calculates addresses and determines which memory is written to; and read operations are only performed for the $\beta$-RAM by the $\beta$-router.

\section{Architecture: Data Processing}
\label{sec:architecture:processor}
As mentioned in Section \ref{sec:algorithm}, our proposed algorithm requires many decoder functions, which translate into instructions that in turn are implemented by specialized hardware blocks.

\begin{figure}
\centering
\begin{tikzpicture}[every text node part/.style={align=center},
  every node/.style={font=\small}]

  \usetikzlibrary{positioning,calc,mux}

  % Define default style for this node
  \tikzset{every mux2 node/.style={draw,minimum width=0.5cm,minimum height=2cm,inner sep=1mm,outer sep=0pt}}
  \tikzset{every mux3 node/.style={draw,minimum width=0.5cm,minimum height=3cm,inner sep=1mm,outer sep=0pt}}
  \tikzset{every mux4 node/.style={draw,minimum width=0.5cm,minimum height=4cm,inner sep=1mm,outer sep=0pt}}

  \tikzset{
    block/.style={draw,rectangle, minimum width = 0.9cm, minimum height = 0.9cm},
    >={latex},
    connector/.style={circle,draw,fill=black,minimum size=3pt,inner sep=0},
  }

  \node[rectangle,inner sep=0pt, minimum width = 3mm] (alpha_in) at (0, 0) {$\mvec{\alpha}$};
  \node[rectangle,inner sep=0pt, minimum width = 3mm] (beta0_in) at (0, 1) {$\mvec{\beta_0}$};
  \node[rectangle,inner sep=0pt, minimum width = 3mm] (zero_in) at (0, 2) {$\mvec{0}$};

  \node[shape=mux2] (m0) at (0.85, 1.5) {$m_0$};

  \draw[-] (zero_in) --  (m0.in0);
  \draw[-] (beta0_in) -- (m0.in1);

  \node[block, minimum height = 2.5cm] (g) at (2, 0.75) {$g$};
  \draw let \p1 = (g.west) in coordinate (g_in0) at (\x1, 1.5);
  \draw let \p1 = (g.west) in coordinate (g_in1) at (\x1, 0);

  \draw[-] (m0.out) -- coordinate (c6a) (g_in0);
  \draw[-] (alpha_in) -- coordinate (c9a) (g_in1) {};

  \node[rectangle,minimum width=0.5] (beta1_in) at (3.25, -0.75) {$\beta_1$};
  \node[block] (spc) at (3.25, 0) {SPC};
  \node[coordinate] at ($(g.east) - (0, 0.75)$) (c3) {};
  \draw[-] (c3) -- coordinate (c4) (spc.west);
  \node[connector] at (c4) {};

  \node[block] (sign) at (3.25, 1) {Sign};
  \draw[-] (c4) |- coordinate (c5) (sign.west);
  \node[connector] at (c5) {};

  \node[block] (ml) at (3.25, 3) {ML};
  \node[block,execute at begin node=\setlength{\baselineskip}{1em}] (repspc) at (3.25, 4) {REP\\SPC};
  \node[block] (rep) at (3.25, 5) {REP};

  \node[shape=mux3] (m3) at (4.5, 0) {$m_3$};
  \node[coordinate, left=0.375cm of m3.in0] (c0) {};
  \draw[-] (sign) -| (c0);
  \draw[-] (c0) -- (m3.in0);
  \draw[-] (spc) -- (m3.in1);
  \node[coordinate, left=0.375cm of m3.in2] (c1) {};
  \draw[-] (beta1_in) -| (c1);
  \draw[-] (c1) -- (m3.in2);

  \node[block, minimum height=3cm] (combine) at (5.5, 1.1) {\rotatebox{90}{COMBINE}};

  \node[coordinate] at ($(combine.west) - (0, 1.1)$) (c2) {};
  \draw[-] (m3.out) -- (c2);

  \node[coordinate, right=0.1cm of c6a] (c6) {};
  \node[coordinate, above=0.625cm of c6] (c7)  {};
  \draw[-] (c6) -- (c7);
  \node[connector] at (c6) {};
  \draw[-] let \p1 = (combine.west), \p2 = (c7) in (c7) -- (\x1, \y2) {};

  \node[block] (f) at (5.5, 7) {$f$};
  \node[coordinate,right= 0.4cm of c9a] (c9) {};
  \node[connector] at (c9) {};
  \draw[-] (c9) |- (f.west);

  \draw let \p1 = (ml.west), \p2 = (c9) in coordinate (c10) at (\x2, \y1);
  \node[connector] at (c10) {};
  \draw[-] (c10) -- (ml.west);

  \draw let \p1 = (repspc.west), \p2 = (c9) in coordinate (c11) at (\x2, \y1);
  \node[connector] at (c11) {};
  \draw[-] (c11) -- (repspc.west);

  \draw let \p1 = (rep.west), \p2 = (c9) in coordinate (c12) at (\x2, \y1);
  \node[connector] at (c12) {};
  \draw[-] (c12) -- (rep.west);

  \node[mux2] (m1) at (7, 6.5) {$m_1$};
  \draw[-] (f.east) -- (m1.in0);
  \draw[-] (c5) |- (m1.in1);
  \node[circle, right=0.5cm of m1.east] (alpha_o) {$\mvec{\alpha'}$};
  \draw[-] (m1.out) -- (alpha_o);

  \node[coordinate, right=0.5cm of combine.east] (c13) {};
  \node[connector] at (c13) {};
  \draw[-] (combine.east) -- (c13);

  \node[mux4] (m2) at (7, 3.4) {$m_2$};
  \node[coordinate, left=2cm of m2.in0] (c14) {};
  \draw[-] (rep.east) -| (c14);
  \draw[-] (c14) -- (m2.in0);
  
  \node[coordinate, left=2cm of m2.in1] (c15) {};
  \draw[-] (repspc.east) -| (c15);
  \draw[-] (c15) -- (m2.in1);
  
  \node[coordinate, left=2cm of m2.in2] (c16) {};
  \draw[-] (ml.east) -| (c16);
  \draw[-] (c16) -- (m2.in2);
  \draw[-] (c13) |- (m2.in3);

  \draw let \p1 = (alpha_o), \p2 = (m2.out) in node[circle] (beta0_o) at (\x1, \y2) {$\mvec{\beta_0'}$};
  \draw[-] (m2.out) -- (beta0_o);

  \draw let \p1 = (alpha_o), \p2 = (c13) in node[circle] (beta1_o) at (\x1, \y2) {$\mvec{\beta_1'}$};
  \draw[-] (c13) -- (beta1_o);

\end{tikzpicture}
\caption{Architecture of the data processing unit.}
\label{fig:processor}
\end{figure}
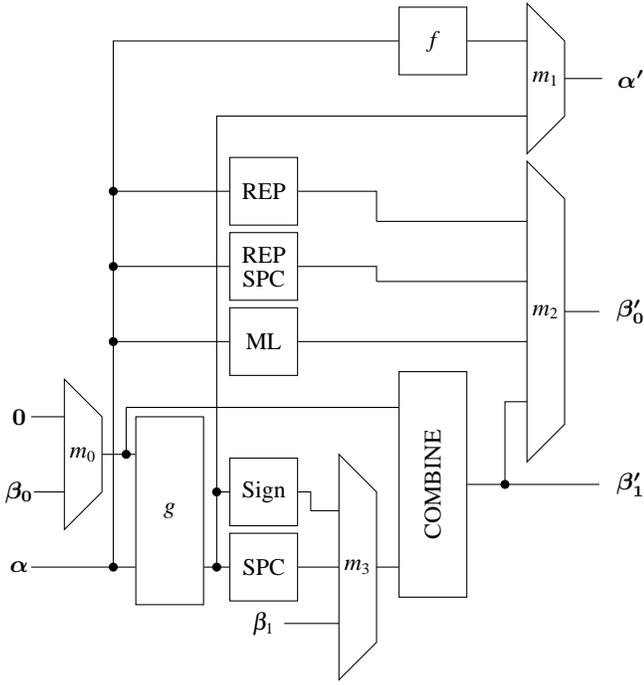

In Fig.~\ref{fig:processor}, which illustrates the architecture of the data processing unit, $\mvec{\alpha}$, $\mvec{\beta}_0$, and $\mvec{\beta}_1$ are the data inputs; while $\mvec{\alpha}'$, $\mvec{\beta}_0'$, and $\mvec{\beta}_1'$ are the corresponding outputs. The first multiplexer ($m_0$) selects either the $\mvec{\beta}_0$ value loaded from memory or the all-zero vector, depending on which opcode is being executed. Another multiplexer ($m_1$) selects the result of $f$ or $g$ as the $\mvec{\alpha}'$ output of the current stage. Similarly, one multiplexer ($m_2$) chooses which function provides the $\mvec{\beta}'_0$ output. Finally, the last multiplexer ($m_3$) selects the input to the COMBINE function.

The critical path of the design passes through $g$, SPC, and COMBINE; therefore, these three blocks must be made fast. As a result, the merged processing element (PE) of \cite{Leroux2013} cannot be used since it has a greater propagation delay than one implementing only $g$. Similarly, using two's complement arithmetic, instead of sign-and-magnitude, results in a faster implementation of the $g$ function as it performs signed addition and subtraction.

In this section, we describe the architecture of the different blocks in detail as well as justify design decisions. We omit the sign block from the detailed description since it simply selects the most significant bit of its input to implement \eqref{eq:threshold}.

\subsection{The $f$ and $g$ Blocks}
As mentioned earlier, due to timing constraints, $f$ and $g$ are implemented separately and use the two's complement representation. The $f$ block contains $P$ $f$ elements which calculate their output by directly implementing \eqref{eq:sc:ms:f}. To simplify the comparison logic, we limit the most negative number to $-2^{Q} + 1$ instead of $-2^{-Q}$ so that the magnitude of an LLR contains only $Q-1$ bits.
The $g$ element also directly implements \eqref{eq:sc:ms:g} with saturation to $2^{Q}-1$ and $-2^{Q} + 1$. This reduction in range did not affect the error-correction performance in our simulations.
The combined resource utilization of an $f$ element and a $g$ element is slightly more than that of the merged PE \cite{Leroux2013}; however the $g$ element is approximately 50\% faster.

Using two's complement arithmetic negatively affected the speed of the $f$ element. This, however, does not impact the overall clock frequency of the decoder since the path in which $f$ is located is short.

Since bit-reversal is used, $f$ and $g$ operate on adjacent values in the input $\mvec{\alpha}$ and the outputs are correctly located in the output $\mvec{\alpha}'$ for all constituent code lengths. Special multiplexing rules would need to be added to support a non-bit-reversed implementation, increasing complexity without any positive effects \cite{Leroux2013}.

\subsection{Repetition Block}
\label{sec:architecture:processor:rep}

The repetition block, described in Section~\ref{sec:algorithm:rep} and denoted REP in Fig.~\ref{fig:processor}, also benefits from using two's complement as its main component is an adder tree that accumulates the input, the sign of whose output is repeated to yield the $\mvec{\beta}$ value. As can be seen in Table~\ref{tab:rep-dist}, the largest constituent repetition code in the polar codes of interest is of length 16. Therefore, the adder tree is arranged into four levels. Since only the sign of the sum is used, the width of the adders was allowed to grow up in the tree to avoid saturation and the associated error-correction performance degradation. This tree is implemented using combinational logic.

When decoding a constituent code whose length $N_v$ is smaller than 16, the last $16 - N_v$ are replaced with zeros and do not affect the result.

An attempt at simplifying logic by using a majority count of the sign of the input values caused significant reduction in error-correction performance that was not accompanied by a perceptible reduction in the resource utilization of the decoder.

\subsection{Repetition-SPC Block}
\label{sec:architecture:processor:rep_spc}
This block corresponds to the very common node with $N_v = 8$ whose left child is a repetition code and its right an SPC code. We implement this block using two SPC nodes and one repetition node. First, four $f$ processing elements in parallel calculate the $\mvec{\alpha}_{REP}$ vector to be fed to a small repetition decoder block. At the same time, both possible vectors of LLR values---$\mvec{\alpha}_{SPC_0}$ and $\mvec{\alpha}_{SPC_1}$, one assuming the output of the repetition code is all zeros and the other all ones---are calculated using eight $g$ processing elements. Those vectors are fed to the two SPC nodes $SPC_0$ and $SPC_1$.

The outputs of these SPC nodes are connected to a multiplexer. The decision $\mvec{\beta}_{REP}$ from the repetition node is used to select between the outputs of $SPC_0$ and $SPC_1$. Finally, results are combined to form the vector of decoded bits $\mvec{\beta}_v$ out of $\mvec{\beta}_{REP}$ and either $\mvec{\beta}_{SPC_0}$ or $\mvec{\beta}_{SPC_1}$. This node is also purely combinational.

\subsection{Single-Parity-Check Block}
\label{sec:architecture:processor:spc}
Due to the large range of constituent code lengths---$[4, 8192]$---that it must decode, the SPC block is the most complex in the decoder. At its core, is a compare-select (CS) tree to find the index of the least reliable input bit as described in Section~\ref{sec:algorithm:spc}. While some small constituent codes can be decoded within a clock cycle; obtaining the input of larger codes requires multiple clock cycles. Therefore, a pipelined design with the ability to select an output from different pipeline stages is required. The depth of this pipeline is selected to optimize the overall decoding throughput by balancing the length of the critical path and the latency of the pipeline.

Table~\ref{tab:spc-dist} was used as the guideline for the pipeline design. As codes with $N_v \in (0, 8]$ are the most common, their output is provided within the same clock-cycle. Using this method, pipeline registers were inserted in the CS tree so that there was a one clock cycle delay for $N_v \in (8, 64]$ and two for $N_v \in (64, 256]$. Since, in the tested codes, SPC nodes only exist in a P-RSPC or a P-0SPC configuration and they receive their input from the $g$ elements, their maximum input size is $P$, not $2P$. Therefore, any constituent SPC code with $N_v > P$ receives its input in multiple clock cycles. The final stage of the pipeline handles this case by comparing the results from the current input word with that of the previous one, and updating a register as required. Therefore, for such cases, the SPC output is ready in $N_v/P + 4$ clock cycles. The extra clock cycle improved operating frequency and the overall throughput.
The pipeline for the parity values utilizes the same structure.

\subsection{Maximum-Likelihood Block}
\label{sec:architecture:processor:ml}
When implementing a length 16 exhaustive-search ML decoder as suggested in \cite{Sarkis2013}, we noted that it formed the critical path and was significantly slower than the other blocks. In addition, once repetition, SPC, and repetition-SPC decoders were introduced, the number of ML nodes of length greater than four became minor. Therefore, the ML node was limited to constituent codes of length four.
When enumerating these codes in the targeted polar codes, we noticed that the one with a generator matrix $G=\left[0 0 0 1; 0 1 0 0\right]$ was the only such code to be decoded with an ML node. The other length-four constituent codes were the rate zero, rate one, repetition, and SPC codes; other patterns never appeared. Thus, instead of implementing a generic ML node that supports all possible constituent codes of length four, only the one corresponding to $G=\left[0 0 0 1; 0 1 0 0 \right]$ is realized. This significantly reduces the implementation complexity of this node.

The ML decoder finds the most likely codeword among the $2^{k_v}=4$ possibilities. As only one constituent code is supported, the possible codewords are known in advance. Four adder trees of depth two calculate the reliability of each potential codeword, feeding their result into a comparator tree also of depth two. The comparison result determines which of $\left[0 0 0 0 \right]$, $\left[0 0 0 1 \right]$, $\left[0 1 0 1 \right]$ or $\left[0 1 0 0 \right]$ is the most likely codeword. This block is implemented using combinational logic only.

\section{Implementation Results}
\label{sec:implementation}

\subsection{Methodology}
The proposed decoder has been validated against a bit-accurate software implementation, using both functional and gate-level simulations. Random test vectors were used. The bit-accurate software implementation was used to estimate the error correction performance of the decoder and to determine acceptable quantization levels.

Logic synthesis, technology mapping, and place and route were performed to target two different FPGAs. The first is the Altera Stratix IV EP4SGX530KH40C2 and the second is the Xilinx Virtex VI XC6VLX550TL-1LFF1759. They were chosen to provide a fair comparison with state of the art decoders in literature. In both cases, we used the tools provided by the vendors, Altera Quartus II 13.0 and Xilinx ISE 13.4. Moreover, we use worst case timing estimates e.g. the maximum frequency reported for the FPGA from Altera Quartus is taken from the results of the ``slow $900\text{mV}$ $85^{\circ}\text{C}$'' timing model.
 
\subsection{Comparison with the State of the Art SC- and SSC-based Polar Decoders}
The fastest SC-based polar decoder in literature was implemented as an application-specific integrated-circuit (ASIC) \cite{Mishra2012} for a (1024, 512) polar code. Since we are interested in better performing longer codes, we compare the proposed decoder with the FPGA-based, length 32768 implementation of \cite{Leroux2013}. Results for the same FPGA are shown in Tables~\ref{tab:impl:sp-sc} and \ref{tab:impl:sp-sc:tp}. For a $(32768, 27568)$ code, our decoder is 15 to 29 times faster than the semi-parallel SC (SP-SC) decoder \cite{Leroux2013}. For the code with a rate of 0.9, it has 19 to 40 times the throughput of SP-SC depending on $P$ and the quantization scheme used, and achieves an information throughput of 1 Gbps for both quantization schemes. It can be also noted that the proposed decoder uses significantly fewer LUTs and registers but requires more RAM, and can be clocked faster. If the decoder followed the buffering scheme of \cite{Leroux2013}, namely, one input frame and no output buffering, its RAM usage would decrease to 507,248 bits for the $P=256$, (7, 5, 1) case and to 410,960 bits when $P = 64$ and the $(6, 4, 0)$ quantization scheme is used.

Although implementation results for $P=256$ are not provided in \cite{Leroux2013}, the throughput the SP-SC algorithm asymptotically approaches $0.5 \cdot f_{clk} \cdot R$ where $f_{clk}$ is the clock frequency. Therefore, even when running at its maximum possible throughput, SP-SC remains 16 to 34 times slower than the proposed decoder for the (32768, 29492) code. The results for the rate 0.9 code with $P = 256$ and the $(7, 5, 1)$ quantization scheme were obtained using Synposis Synplify Premier F-2011.09-SP1-1 and Altera Quartus 11.1.

\begin{table}[t]
  \centering
  \caption{Post-fitting results for a code of length 32768 on the Altera Stratix IV EP4SGX530KH40C2.}
  \begin{tabular}{c c c D{,}{,}{1.2} D{,}{,}{1.2} D{,}{,}{1.1} c }
    \toprule
    \multirow{2}{*}{Algorithm} & \multirow{2}{*}{P} & \multirow{2}{*}{Q} & \multicolumn{1}{c}{\multirow{2}{*}{LUTs}} & \multicolumn{1}{c}{\multirow{2}{*}{Registers}} & \multicolumn{1}{c}{RAM} & $f$ \\
    & & & & & \multicolumn{1}{c}{(bits)} & (MHz)\\
    \midrule
    SP-SC\cite{Leroux2013} & 64  & 5         & 58,480 & 33,451 & 364,288 &  66\vspace{4pt}\\
    This work              & 64  & $(6,4,0)$ &  6,830 &  1,388 & 571,800 & 108 \\
                           &     & $(7,5,1)$ &  8,234 &  \multicolumn{1}{c}{$\quad\;$858} & 675,864 & 100\vspace{2pt}\\
                           & 256 & $(6,4,0)$ & 25,866 &  7,209 & 536,136 & 108\\
                           &     & $(7,5,1)$ & 30,051 &  3,692 & 700,892 &  104 \\

    \bottomrule
  \end{tabular}
  \label{tab:impl:sp-sc}
\end{table}

\begin{table}
  \centering
  \caption{Information throughput comparison for codes of length 32768 on the Altera Stratix IV EP4SGX530KH40C2.}
  \begin{tabular}{c c c c c}
    \toprule
    Algorithm & \multicolumn{1}{c}{Code rate} & P & Q & \multicolumn{1}{c}{T/P (Mbps)} \\
    \midrule
    SP-SC\cite{Leroux2013} & 0.84 & 64  & 5         & 26 \\
                           & 0.9  & 64  & 5         & 28\vspace{4pt}\\
    This work              & 0.84 & 64  & $(6,4,0)$ & 425\\
                           &      &     & $(7,5,1)$ & 406\vspace{1pt}\\
                           &      & 256 & $(6,4,0)$ & 791 \\
                           &      &     & $(7,5,1)$ & 775\vspace{2pt}\\
                           & 0.9  & 64  & $(6,4,0)$ & 547\\
                           &      &     & $(7,5,1)$ & 523\vspace{1pt}\\
                           &      & 256 & $(6,4,0)$ & 1,081\\
                           &      &     & $(7,5,1)$ & 1,077\\
    \bottomrule
  \end{tabular}
  \label{tab:impl:sp-sc:tp}
\end{table}

The two-phase successive-cancellation (TPSC) decoder is an SC-based decoder that optimizes the algorithm to reduce memory \cite{Pamuk2013} and employs elements of SSC decoding to improve throughput. It is limited to values of $N$ that are even-powers of two. Therefore, in Table~\ref{tab:impl:ssc} we utilize a (16384, 14746) code constructed for $E_b/N_0 = 5$ dB and compare the resulting resource utilization and information throughput with the results of \cite{Pamuk2013}. The quantization schemes used were $(6, 4, 0)$ for the proposed decoder and 5 bits for TPSC. Since \cite{Pamuk2013} does not include the input buffers necessary to sustain the presented throughput, Table~\ref{tab:impl:ssc} provides an extra entry, denoted TPSC*, that includes the added RAM required to buffer a second input frame. From the table, it can be observed that the proposed algorithm is eight times faster than TPSC even though the latter is running at more than twice the frequency. Additionally the proposed algorithm uses 1.7 times the LUTs and 1.2 times the registers of TPSC. When both decoder include buffers to store two received frames, the proposed algorithm uses 1.4 times the RAM of TPSC. Based on this comparison, it can be concluded that TPSC cannot match the throughput of the proposed algorithm with the same complexity by utilizing multiple decoders decoding different frames simultaneously since the resulting TPSC system will utilize more than four times the resources of the proposed decoder. The last entry in the table presents the results achievable by the proposed decoder with $P = 256$, where the information throughput is $\sim 1.1$ Gbps.

\begin{table}[t]
  \centering
  \caption{Post-fitting and information throughput results for a (16384, 14746) code on the Altera Stratix IV EP4SGX530KH40C2.}
  \begin{tabular}{c c D{,}{,}{1.2} D{,}{,}{1.2} D{,}{,}{1.1} c c}
    \toprule
    \multirow{2}{*}{Algorithm} & \multirow{2}{*}{P} &  \multicolumn{1}{c}{\multirow{2}{*}{LUTs}} & \multicolumn{1}{c}{\multirow{2}{*}{Reg.}} & \multicolumn{1}{c}{RAM} & $f$ & T/P\\
    & & & & \multicolumn{1}{c}{\footnotesize(bits)} & {\footnotesize(MHz)} & {\footnotesize(Mbps)}\\
    \midrule
    TPSC\cite{Pamuk2013} & 128 & 7,815 & 3,006 & 114,560 & 230 & 106\\
    TPSC*\cite{Pamuk2013} & 128 & 7,815 & 3,006 & 196,480 & 230 & 106\\
    This work & 128 & 13,388 & 3,688 & 273,740 & 106 & 824\\
    This work & 256 & 25,219 & 6,529 & 285,336 & 106 & 1,091\\
    \bottomrule
  \end{tabular}
  \label{tab:impl:ssc}
\end{table}

\subsection{Comparison with an LDPC code of similar error correcting performance}
A fully-parallel (2048, 1723) LDPC decoder on FPGA is presented in \cite{Torres2012}. At 30.7~MHz on a Xilinx Virtex VI XC6VLX550TL, an information throughput of 1.1~Gbps is reached. Early termination could be used to achieve 8.8~Gbps at 5~dB, however that would require support for early termination circuitry and extra buffering that were not implemented in \cite{Torres2012}.

\begin{table}
  \centering
  \caption{Comparison with an LDPC code of similar error correcting performance, on the Xilinx Virtex VI XC6VLX550TL.}
  \begin{tabular}{c c D{,}{,}{1.2} D{.}{.}{1.1} D{.}{.}{1.3} }
    \toprule
    Code                  & Q & \multicolumn{1}{c}{LUTs} & \multicolumn{1}{c}{$f_{max}$ (MHz)} & \multicolumn{1}{c}{T/P (Gbps)} \\
    \midrule
    LDPC\cite{Torres2012} & 4         & 99,468 & 30.7 & 1.102 \\
    This work             & $(6,4,0)$ & 18,024 & 71.3 & 0.542 \\ %% 71.327 MHz
                          & $(7,5,1)$ & 21,700 & 71.0 & 0.539 \\ %% 70.977 MHz
    \bottomrule
  \end{tabular}
  \label{tab:impl:ldpc}
\end{table}

Results for our decoder with $P=256$ and a (32768, 27568) polar code implemented on the same FPGA as the LDPC decoder are shown in Table~\ref{tab:impl:ldpc}. Our decoder requires 5 times fewer LUTs, but only achieves half of the throughput.

\section{Conclusion}
\label{sec:conclusion}
In this work we presented a new algorithm for decoding polar codes that results in a high-throughput, flexible decoder. An FPGA implementation of the proposed algorithm was able to achieve an information throughput of 1 Gbps when decoding a (32768, 29492) polar code with a clock frequency of 108 MHz. We expect derivative works implementing this decoder as an ASIC to reach a throughput of 3 Gbps when operating at 300 MHz with a complexity lower than that required by LDPC decoders of similar error correction performance. Thus, our results indicate that polar codes are promising candidates for data storage systems.

\section*{Acknowledgement}
The authors wish to thank CMC Microsystems for providing access to the Altera, Xilinx, Mentor Graphics and Synopsys tools.
The authors would also like to thank Prof. Roni Khazaka and Alexandre Raymond of McGill University for helpful discussions.
Claude Thibeault is a member of ReSMiQ.

\bibliographystyle{IEEEtran}
\bibliography{IEEEabrv,polar-up.bib}

\end{document}